\documentclass[apj]{emulateapj} 
\usepackage{color}
\usepackage{hyperref} 

\shorttitle{The SDSS Coadd: Cosmic Shear}
\shortauthors{Lin et al.}

\newcommand{\sfig}[2]{
\includegraphics[width=#2]{#1}
        }
\newcommand{\Sfig}[2]{
    \begin{figure}[h]
    \sfig{#1.eps}{1.0\columnwidth}
    \caption{{\small #2}}
    \label{fig:#1}
    \end{figure}
}

\newcommand{\SfigT}[2]{
    \begin{figure*}[thbp]
    \sfig{#1.eps}{2.0\columnwidth}
    \caption{{\small #2}}
    \label{fig:#1}
    \end{figure*}
}
\newcommand{\Sfigtwo}[3]{
	\begin{figure*}[thbp]
\sfig{#1.eps}{1.0\columnwidth}
\sfig{#2.eps}{1.0\columnwidth}
\caption{{\small #3}}
\label{fig:#1}
\end{figure*}
}
\newcommand{\Sfigfour}[5]{
	\begin{figure*}[thbp]
\sfig{#1.eps}{1.05\columnwidth}
\sfig{#2.eps}{1.05\columnwidth}
\sfig{#3.eps}{1.05\columnwidth}
\sfig{#4.eps}{1.05\columnwidth}
\caption{{\small #5}}
\label{fig:#1}
\end{figure*}
}

\newcommand{\rf}[1]{\ref{fig:#1}}

\def\be{\begin{equation}}
\def\ee{\end{equation}}
\def\bea{\begin{eqnarray}}
\def\eea{\end{eqnarray}}
\newcommand{\vs}{\nonumber\\}

\newcommand{\Cee}{C_{EE}}
\newcommand{\Cbb}{C_{BB}}
\newcommand{\Ceb}{C_{EB}}
\newcommand{\Cov}{{\rm Cov}}
\newcommand{\tilga}{\tilde \gamma}

\newcommand{\ec}[1]{Eq.~(\ref{eq:#1})}
\newcommand{\Ec}[1]{(\ref{eq:#1})}
\newcommand{\eql}[1]{\label{eq:#1}}
\newcommand{\gam}[2]{\gamma_{#1}(#2)}
\newcommand{\vecn}[1]{\vec{n}_{#1}}

\begin{document}

\title{The SDSS Coadd: Cosmic Shear Measurement}         

\author{
Huan Lin$^{1}$,
Scott Dodelson$^{1,2,3}$,
Hee-Jong Seo$^{4}$,
Marcelle Soares-Santos$^{1}$,
James Annis$^{1}$,
Jiangang Hao$^{1}$,
David Johnston$^{1}$,
Jeffrey M. Kubo$^{1}$,
Ribamar R. R. Reis$^{1,5}$,
Melanie Simet$^{2,3}$
}

\affil{
${}^{1}$Center for Particle Astrophysics, Fermi National Accelerator Laboratory, Batavia, IL 60510 \\
${}^{2}$Department of Astronomy \& Astrophysics,
The University of Chicago, Chicago, IL 60637 \\
${}^{3}$Kavli Institute for Cosmological Physics, Chicago, IL 60637 \\
${}^{4}$Berkeley Center for Cosmological Physics, LBL and Department of
Physics, University of California, Berkeley, CA, USA 94720 \\
${}^{5}$Instituto de F\'\i sica, Universidade Federal do Rio de Janeiro, CEP 21941-972, Rio de Janeiro, RJ, Brazil \\
}

\begin{abstract}
Stripe 82 in the Sloan Digital Sky Survey was observed multiple times, allowing deeper images to be constructed
by coadding the data. Here we analyze the ellipticities of background galaxies in this 275 square degree region, searching
for evidence of distortions due to cosmic shear. The E-mode is detected 
in both real and Fourier space with $>5$-$\sigma$
significance on degree
scales, while the B-mode is consistent with zero as expected. 
The amplitude 
of the signal constrains the combination of the matter density $\Omega_m$ and fluctuation amplitude $\sigma_8$ to be $\Omega_m^{0.7}\sigma_8 = 0.252^{+0.032}_{-0.052}$.
\end{abstract}

\keywords{cosmological parameters --- cosmology: observations ---
gravitational lensing --- large-scale structure of universe
}

\section{Introduction}

Since the first detections of cosmic shear \citep{VanWaerbeke:2000,Bacon:2000,Wittman:2000},
gravitational lensing has emerged as a powerful tool in the quest 
to pin down cosmological parameters 
\citep{Rhodes:2001,vanWaerbeke:2001,Hoekstra:2002,Jarvis:2003,Bacon:2003,Haman:2003,%
Heymans:2004,Rhodes:2004,Fu:2007qq,Schrabback:2010}. 
Distortions in the shapes of distant
galaxies depend on the intervening cosmic shear field, and careful 
observations of the ellipticities of many background galaxies enable us
to measure the statistics of this field, and compare with the predictions of a
given cosmological model. This comparison is most  robust on large
scales, which are unaffected by nonlinearities and baryonic effects. 
However, observations are easiest on the small scales covered by deep surveys 
where the point spread function is relatively stable, 
so only recently have detections moved to larger scales. 
Indeed, one of the most promising 
applications of weak lensing is to measure properties of dark energy
\citep[see][for a review]{Munshi:2008}, and 
for this purpose, precise measurements on scales of order ten arcminutes and 
larger will be most constraining. 

With this in mind, we have measured the ellipticities of galaxies in Stripe 82 of the Sloan Digital Sky 
Survey \citep[SDSS;][]{York:2000}, 
a rectangular ($2.5^\circ\times110^\circ$) region on the sky that was imaged 
multiple times. The images have been coadded leading to a much deeper picture 
of the galaxies in the Universe than is available from the part of the survey 
comprised of single images \citep{Annis:2011}. 
The relatively large area and deep images offer a 
glimpse into the future, as large scale surveys such as the 
Dark Energy Survey \citep[DES;][]{Abbott:2005bi} and the 
Large Synoptic Survey Telescope \citep[LSST;][]{lsst} come on line. As we 
present results obtained on the coadded data, an important consideration is
the systematics associated with the coadd. How careful must one be when 
combining multiple images of the sky? 

Section~\ref{sec:data} describes the data set, the coaddition method, the 
correction for the effects of the point spread function (PSF) modeling, and 
the prescription for obtaining photometric redshifts. 
Section~\ref{sec:theory} reviews the different two-point functions used to 
characterize the shear distribution and how these are related to the 
underlying matter power spectrum. Section~\ref{sec:xi} 
presents the correlation function results for both mock catalogs (to obtain a 
benchmark against which the actual data can be compared) and the data. 
Section~\ref{sec:power} 
presents a complementary approach by estimating the Fourier space power spectrum in several 
different ways. In each case (real space with the correlation function and 
Fourier space with the power spectrum), we isolate modes that should be 
non-zero and modes that arise due to systematics and show that the former 
are detected and the latter are consistent with zero. Finally, in 
Section~\ref{sec:cosmo} we use the 
two-point function results to obtain constraints on the fluctuation amplitude 
$\sigma_8$ and matter density $\Omega_m$. Section~\ref{sec:conclusions}
summarizes our results and conclusions.

While this work was underway, we learned of a parallel effort by 
\cite{Huff:2011}.  
These two efforts use different methods of coaddition and different sets of 
cuts for the input images and galaxies; what they have in common is their use 
of SDSS data (not necessarily the same set of runs) and their use of the 
SDSS {\tt PHOTO} pipeline for the initial reduction of the single epoch data 
and the final reduction of the coadded data (however, they use different 
versions of {\tt PHOTO}).  Using these different methods, both groups have 
attempted to extract the cosmic shear signal and its cosmological 
interpretations.
We have coordinated submission with them but have not consulted their results 
prior to this, so these two analysis efforts are completely independent, 
representing an extreme version of two independent pipelines.

\section{Stripe 82 Coadd Data}\label{sec:data}
\subsection{Coadd}
The SDSS \citep{York:2000} obtained CCD imaging in five bands over 10,000 
square degrees in the Northern Galactic Cap. In addition, the SDSS imaged 
the Celestial Equator in the Southern Galactic Cap (Stripe 82) multiple times 
during the Fall months when the Northern Cap was not observable. 
The SDSS Coadd \citep{Annis:2011} is a 275 square degree survey resulting from 
the stacking of  20-30 exposures on Stripe 82. The depth achieved is  
2 magnitudes fainter and the seeing is 0.3\arcsec\ better  
than the SDSS Northern Cap data, which is comprised of 
single exposures. 
\cite{Annis:2011} provides a detailed description of the coadd construction 
process and the resulting data set, 
which includes images and catalogs, all publicly available as part of the 
SDSS Data Release 7  \citep{2009ApJS..182..543A}. 
Here we summarize the relevant aspects for this work.

SDSS imaging is obtained in a time-delay-and-integrate (or drift scan) mode 
in five filters $ugriz$ \citep{Fukugita:1996} using the 2.5 degree wide SDSS 
imaging camera \citep{Gunn:1998}. The coadd area is, therefore, 2.5 degree 
wide and covers the range $-50 \le {\rm RA} \le 60$ degrees on Stripe 82.
For most of the program images were taken under photometric conditions and 
good seeing, but the SDSS Supernova program took data in the same area
even under non-photometric conditions when the seeing was poor. 

Although Stripe 82 has been imaged more than 100 times, the number of 
exposures  included in the coadd is 20-30 because the data 
were selected using various quality criteria and, at the time
of processing, only data up to December 1 2005 were available. 
Fields were selected based on $r$-band parameters. 
In addition to the basic 
requirement that the fields contain enough stars for relative calibration, 
the selection allowed at the most 
$ 2$\arcsec\ PSF FWHM, 0.5 mag of sky noise increase and  0.2 mag of extinction.  
Rejection of entire fields 
based on $r$-band parameters  maximizes the homogeneity 
of the input data for the coadd construction.

The selected fields undergo photometric calibration and sky subtraction. Masks accounting for
Stripe 82 geometry and bad/saturated pixels are created as well as the inverse variance and weight maps.  All  images, maps and masks are aligned on a rectangular grid with the appropriate dimensions  
($-50^\circ \le {\rm RA} \le 60^\circ$ and $-1.25^\circ \le {\rm Dec} \le 1.25^\circ$)
and in the usual SDSS image format (1489 rows along ${\rm RA}$ and 2048 
columns along ${\rm Dec}$, at $0.396''/{\rm pix}$ scale in a J2000 coordinate system).

The coaddition itself is done using a weighted clipped mean on an image by image
basis where $w_i=T_i/(\sigma_i \times {\rm FWHM_i})^{2}$ is the weight and 
$T_i$, $\sigma_i$ and FWHM$_i$ are the sky transparency, sky noise and seeing of the $i^{th}$ image.
Good seeing data taken when the 
sky is clear and the atmospheric glow is at a minimum are weighted higher by 
this scheme, which reduces the average PSF in the coadd images to $1.1''$ 
(median seeing for the SDSS single exposure data is $1.4''$).
The PSFs in the Coadd were obtained by adding the input PSFs using the 
same weights used for the images. The resulting PSF was used as input to 
the SDSS photometric pipeline ({\tt PHOTO}, \cite{Lupton:2001}) which produces the galaxy catalog 
used as base for this work. The quantities listed in the catalog are the same 
quantities reported for the SDSS main survey catalog, including measurements 
of the second and fourth moments for each galaxy and of the PSF at the position of each galaxy. 

\subsection{Photometric redshifts}\label{sec:photoz}
Photometric redshifts are crucial for this work, as well as related projects.
A neural network algorithm that was successfully used 
in the SDSS DR6 \citep{Oyaizu:2007jw} was applied to the coadd galaxies.
The resulting photometric redshift galaxy catalog is fully described in 
\cite{Reis:2011} and is publicly available as an SDSS DR7 value-added catalog. 
Here we present an overview of the
method and the catalog properties that are relevant for this work.

We use a particular type of Adaptive Neural Network called Feed Forward Multilayer
Perceptron (FFMP) to map the relationship between photometric observables
and redshifts \citep[for details see][]{Reis:2011,Oyaizu:2007jw}.
An FFMP network consists of several input nodes, one or more hidden layers,
and several output nodes, all interconnected by weighted connections.
Once the network configuration is specified, it can be trained to
output an estimate of redshift given the input photometric observables.
The training process involves
finding the set of weights that gives the best photometric redshift estimate
for the training set (a sample of galaxies with spectroscopic redshifts). 
These weights are then applied to the full photometric 
sample to produce a photometric redshift catalog. Errors are estimated
using the Nearest Neighbor Error estimator \citep{Oyaizu:2007kd}. 
This estimator
associates photo-$z$ errors to photometric objects by considering the
errors for objects with similar multi-band magnitudes in the
validation set (a second sample of galaxies with spectroscopic redshifts).

A spectroscopic sample of $82,741$ galaxies was established by gathering 
data from various surveys overlapping Stripe 82. 
69\% of the sample was obtained from the SDSS data  \citep{2009ApJS..182..543A}, 
12\% from DEEP2 \citep{Weiner:2004qm},
11\% from WiggleZ \citep{Drinkwater:2009sd},
 7\% from VVDS \citep{2008A&A...486..683G},
  and
 2\% from CNOC2 \citep{Yee:2000qj}.
The full sample is divided in two sets of equal size, for training and 
validation respectively. The best results are obtained when 
magnitudes are used as input parameters and the training is performed in 
independent magnitude slices in the $r$-band. 

The resulting galaxy catalog, with photometric redshift measurements, 
forms the base catalog for this work.
The photometric redshifts are well measured up to $z\sim0.8$,
the mean photo-$z$ error of the validation set galaxies is $\sigma_z = 0.031$,
and the average estimated photo-$z$ error for the full sample is 
$\sigma_z = 0.18$.
Our analysis relies on the redshift probability distribution function of the galaxies 
used in the cosmic shear measurement. As discussed in the following section, we perform several cuts in 
the galaxy catalog to mitigate the systematic effects. We cut in $i$-band magnitude ($18<i<24$), size ($>1.5 \times$ the PSF size) and ellipticity components ($<1.4$).  
The overall photometric redshift distribution, for the galaxies remaining after these cuts 
is shown in Fig. \rf{dndz} (grey histogram).
Motivated by the fact that the photometric redshift distribution is also a source of systematics 
in our analyses, and that we observe a narrow peak in the distribution at $z\sim0.8$ due to large
errors  at high $z$, we tested two cuts in photometric redshift errors, $\sigma_{z}<0.20$ and 
$\sigma_{z}<0.15$. These are also shown in 
Fig. \rf{dndz} (blue and red histograms, respectively) and the corresponding results 
will be discussed in the following  sections. 
\begin{figure}[!htb]
\centering
\includegraphics[width=1.\columnwidth]{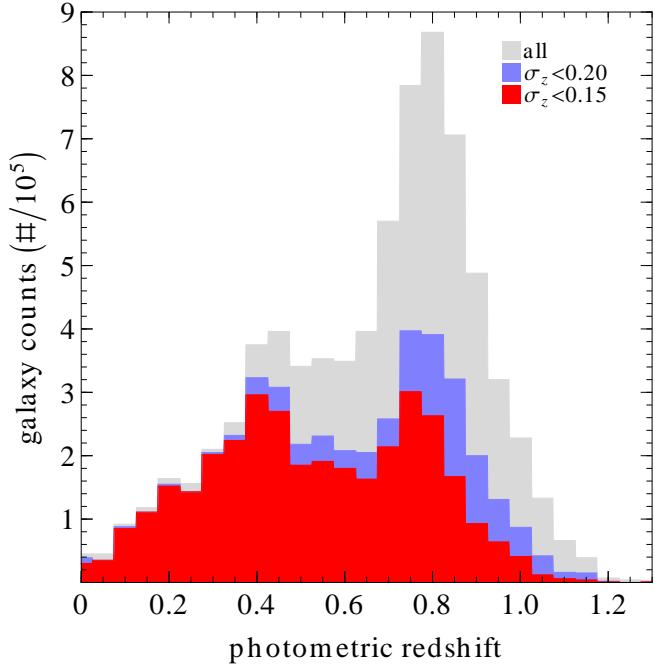}
\caption[Photoz distribution.]{Photometric redshift  distribution for galaxies passing the magnitude,
size and ellipticity cuts (grey) and the photometric redshift error cuts $\sigma_{z}<0.20$ 
(blue) and $\sigma_{z}<0.15$ (red).}
\label{fig:dndz}
\end{figure}

\subsection{Shape Measurement and PSF Correction}\label{sec:shape_psf}

The initial shape measurement is performed using the SDSS {\tt PHOTO} pipeline 
\citep{Lupton:2001}, which measures the shapes of objects using 
adaptive moments \citep{Bernstein:2002}.  However,
we find that we need to correct the point spread function (PSF) model computed
by the SDSS reduction pipeline, as there are small but significant
systematic offsets between the PSF model adaptive moments and the 
same quantities as directly measured on bright stars.
For example, the top panel of Fig.~\rf{e1_residuals} shows the residuals in
the ellipticity component $e_1$, relative to the PSF model, for unsaturated
bright stars with $16 < i < 17$.  
Note that our $e_1, e_2$ convention is such that the positive $x$ and $y$ 
directions are aligned along the positive Dec and RA directions, respectively.
The $e_1$ residuals show a small overall bias ($\Delta e_1 = 0.004$), 
as well as conspicuous trends ($\Delta e_1 \sim 0.01$)
as a function of declination, plus discontinuities between
neighboring CCD camera columns (``camcols'').  Note there are 12 camcols 
in all:  6 physical columns of CCDs in the SDSS imaging camera, which are then
interleaved by the North and South scans of Stripe 82.  In order to remove 
these PSF residuals, which would otherwise contribute systematic errors to 
the cosmic shear measurements,  we make corrections to the PSF model by 
fitting polynomials to the residuals along the declination direction, 
separately for each of the 12 camcols, and for each of the 4 adaptive 
moments quantities relevant to the linear PSF correction scheme we will
use \citep[Appendix B of][]{Hirata:2003}.  Specifically,
we fit quadratic polynomials in declination to the residuals in the
two ellipticity components ({\tt mE1-mE1PSF}, {\tt mE2-mE2PSF}) and in the size
({\tt mRrCc - mRrCcPSF}), and we fit linear polynomials to the residuals in 
the 4th moment ({\tt mCr4 - mCr4PSF}).  The bottom panel of 
Fig.~\rf{e1_residuals} shows the improved $e_1$ residuals after applying the 
correction procedure of subtracting off the best-fit polynomial to the 
original $e_1$ residuals vs.\ declination.  The improvement in the 
residuals is visible, for example, at declination~$\approx -1^\circ$, at the boundary 
between the first two camcols: the discontinuity visible there in the top 
panel is gone in the bottom panel, after the application of our correction 
procedure.

\Sfig{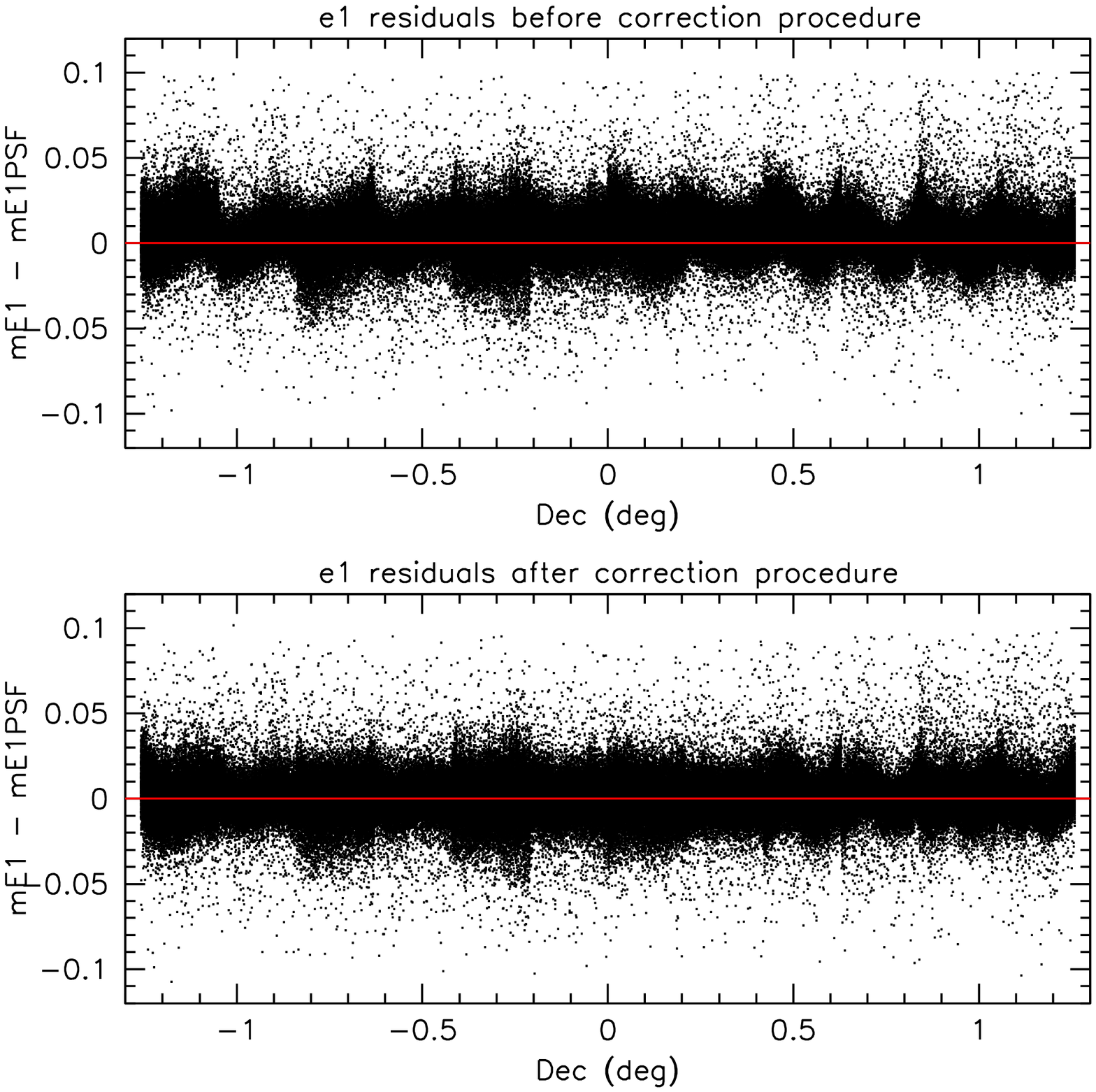}{{\it Top panel:} The difference between the measured
ellipticity component $e_1$ and the PSF model $e_1$, for unsaturated bright
stars with $16 < i < 17$, plotted against declination.  Note the small overall
bias and the trends and discontinuities in these $e_1$ residuals.
{\it Bottom panel:} Same as the top but after applying the PSF correction 
procedure described in the text, which removes the bias, as well as
reduces the trends and discontinuities in the $e_1$ residuals.}

After correcting the PSF model as described above, we proceed to 
calculate PSF-corrected galaxy ellipticities, using
the linear PSF correction algorithm described in \citet{Hirata:2003}.  
The galaxies used in our lensing analysis are required to be 
classified by {\tt PHOTO} as galaxies (type=3), 
have extinction corrected model magnitudes \citep{Stoughton:2002} in the range
$18 < i < 24$, not contain saturated pixels, and not have flags indicating
problems with the adaptive moments measurements.
We use only the shape measurements from the $i$ band as it is the filter that 
has the best seeing ($1.05''$) in the coadd, as shown in \citet{Annis:2011}.
Similar to \citet{Mandelbaum:2005},
we use the {\tt PHOTO} star/galaxy classification and restrict the sample to galaxies at least 50\% larger than the PSF. This is quantified by requiring the {\it resolution factor} $\rm{R}>0.33$, where  
R (not to be confused with responsivity $R$ introduced below) is given by
\begin{equation}
\mathrm{R}=1-\frac{M^{\mathrm{PSF}}_{\mathrm{rrcc}}}{M_{\mathrm{rrcc}}}
\end{equation}
where $M_{\rm{rrcc}}$ and $M_{\rm{rrcc}}^{\rm{PSF}}$ are the sum of the second order moments (in the CCD row and column directions) of the object and PSF 
respectively.
Similar to the SDSS lensing analysis of \citet{Hirata:2004}, we further 
restrict the galaxies used in our study to those with PSF-corrected 
ellipticities $|e_1| < 1.4$ and $|e_2| < 1.4$.  
The number of galaxies used in our analysis is 
3.70 million for the photo-$z$ error $\sigma_z < 0.15$ sample and 
4.69 million for the $\sigma_z < 0.2$ sample.  
These numbers correspond to surface densities of 3.7 and 4.7 galaxies
per arcmin$^2$, respectively.
The rms ellipticity
for our galaxies is $\sigma_e = 0.44$ (for each of the $e_1$ and $e_2$ 
components) for the $\sigma_z < 0.15$ sample and $\sigma_e = 0.47$
for the $\sigma_z < 0.2$ sample.  Note that $\sigma_e^2$ is effectively
the average value of the quadrature sum of the intrinsic shape noise and 
the ellipticity measurement error (and a small contribution from the cosmic shear signal) 
for each galaxy.  Following
\citet{Hirata:2004}, the intrinsic shape noise contribution alone is
$e_{\rm int} = 0.37$ per ellipticity component, resulting in a shear 
responsivity $R = 2(1-e_{\rm int}^2) = 1.7$.  We then convert from
ellipticity to shear $\gamma$ using $\gamma = e / R$.

We have found that the PSF-corrected galaxy ellipticities $e_1$ and $e_2$ 
have typical average values over each CCD camcol of 
$|\bar{e}_1|$, $|\bar{e}_2| \approx 0.003$, whereas we would have expected 
that the average over a thin but very long $\sim$0.2$^\circ$ by 110$^\circ$ 
camcol should be closer to zero, i.e., $|\bar{e}| \approx 8 \times 10^{-4}$, 
given the measured rms ellipticity $\sigma_e = 0.45$ and the average of 
3.5$\times 10^5$ galaxies per camcol.  We have attributed this additive bias
in the ellipticity measurements to remaining camcol-dependent systematic errors
in our PSF model, and have chosen to correct for the bias by subtracting 
off the mean of the galaxy $e_1$ and $e_2$ values on a camcol-by-camcol basis.
If not corrected, the result is to cause additive offsets in the 
correlation function $\xi$ of order
$|\Delta\xi| \sim \bar{e}^2 / R^2 \sim 3\times 10^{-6}$, comparable to the 
cosmic shear signal we will measure in Section~\ref{sec:xi_data}.
In Section~\ref{sec:systematics} we will assess the potential impact on our 
lensing analysis of any residual systematic effects due to the PSF, 
by checking the cross correlation function of the PSF and galaxy ellipticities.

\Sfigtwo{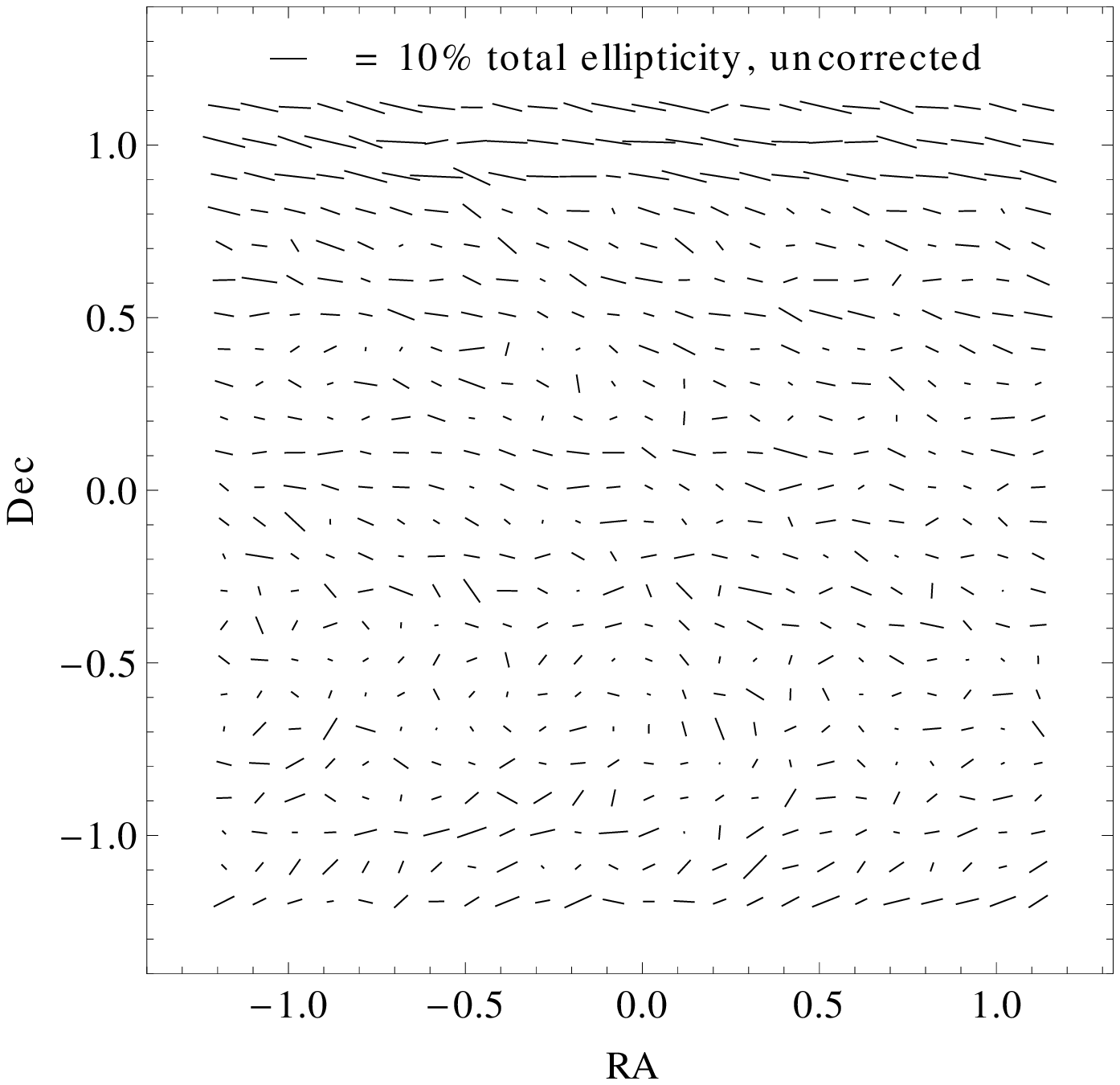}{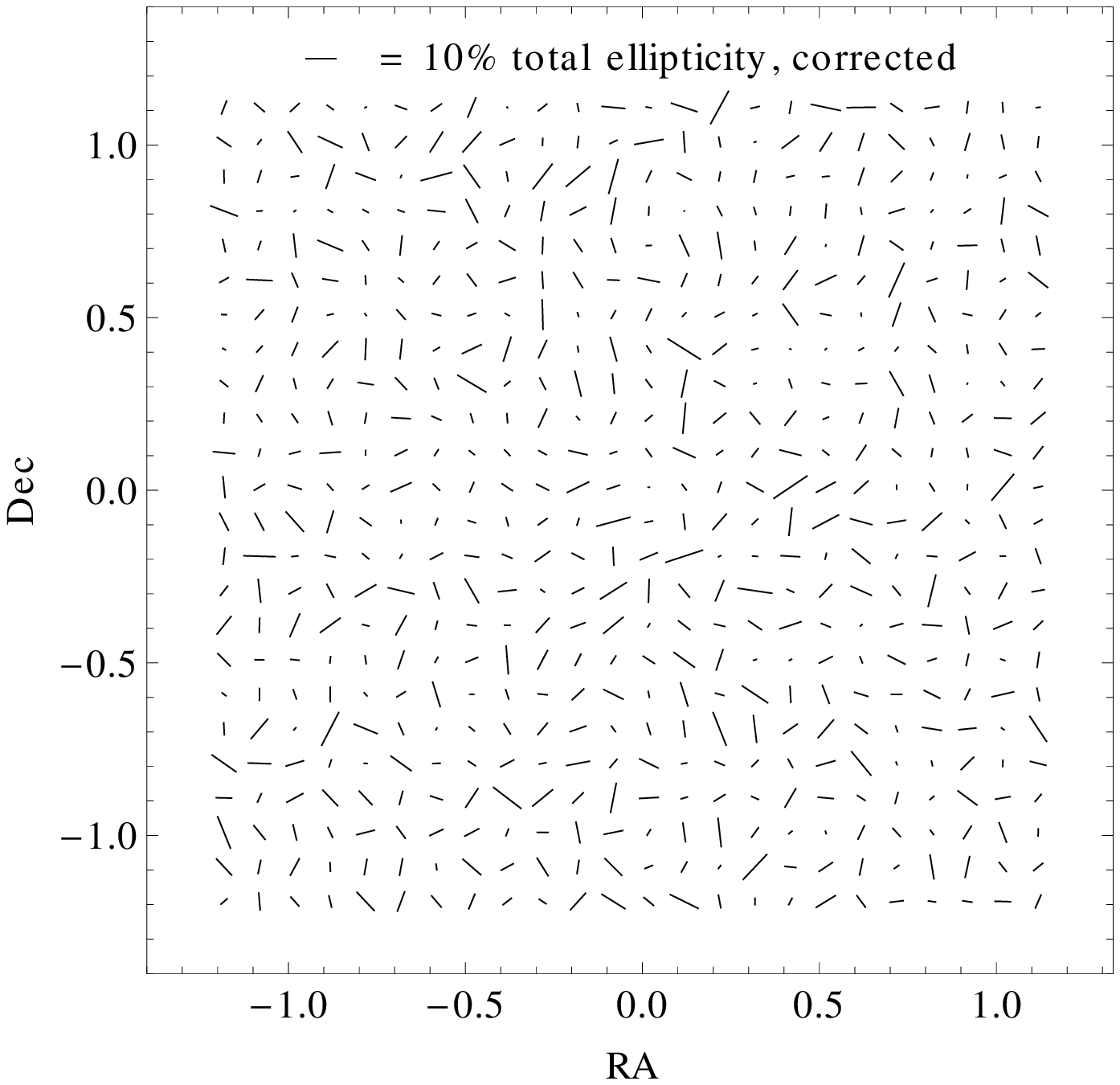}{Map of the galaxy ellipticities,
averaged over the $0.1^\circ \times 0.1^\circ$ pixels used in our analysis, for a
small region of the Stripe 82 area.  The map is shown both before (left) and 
after (right) PSF correction, showing the effective removal of the PSF-induced
ellipticities seen in particular along the top of the map on the left.  
For reference, a stick of ellipticity $e = 0.1$ is labeled at the top of 
each panel.}

Next, we divide the Stripe 82 area into square pixels of size 
$0.1^\circ \times 0.1^\circ$, and compute the average PSF-corrected 
ellipticity components $e_1$ and $e_2$ in each pixel for all the galaxies 
meeting our lensing sample criteria.  
We use these averaged ellipticity values in our subsequent analyses.
Fig.~\rf{shear_before} shows an example map of galaxy ellipticities
for a small part of the Stripe 82 area.  The map is shown both before and 
after PSF correction, in particular illustrating the efficacy in removing the 
PSF-induced ellipticity patterns in the outer CCD camera columns, along the 
top of Fig.~\rf{shear_before}.  The distributions over our full
data area of the pixel-averaged $e_1$ and $e_2$ values are plotted in
Fig.~\rf{ecut14_photozerr15_binned_e1orig_noboss_nobias}, both before and after PSF correction.
The pre-corrected galaxy ellipticity distributions clearly show the effects 
of the PSF.  In particular, the main feature is the negative mean and tail seen
in the $e_1$ distribution, indicating an elongation of the galaxy shapes along
the RA direction, which is also the scan direction of the coadd imaging runs.
In contrast, after applying the PSF correction procedure as described above,
the galaxy ellipticity distributions are seen to be much better behaved,
without any conspicuous asymmetry or bias.  In fact, the post-correction
$e_1$ and $e_2$ distributions are each well approximated by a Gaussian with 
$\sigma \approx 0.04$.  This value of $\sigma$ is consistent with the 
single-galaxy rms ellipticity, $\sigma_e = 0.45$, divided by the square root 
of the average number of galaxies per pixel $\bar{N}$, 
where $\bar{N} = 129$ and 163 for the $\sigma_z < 0.15$ and 0.20 samples, 
respectively.

\Sfigfour{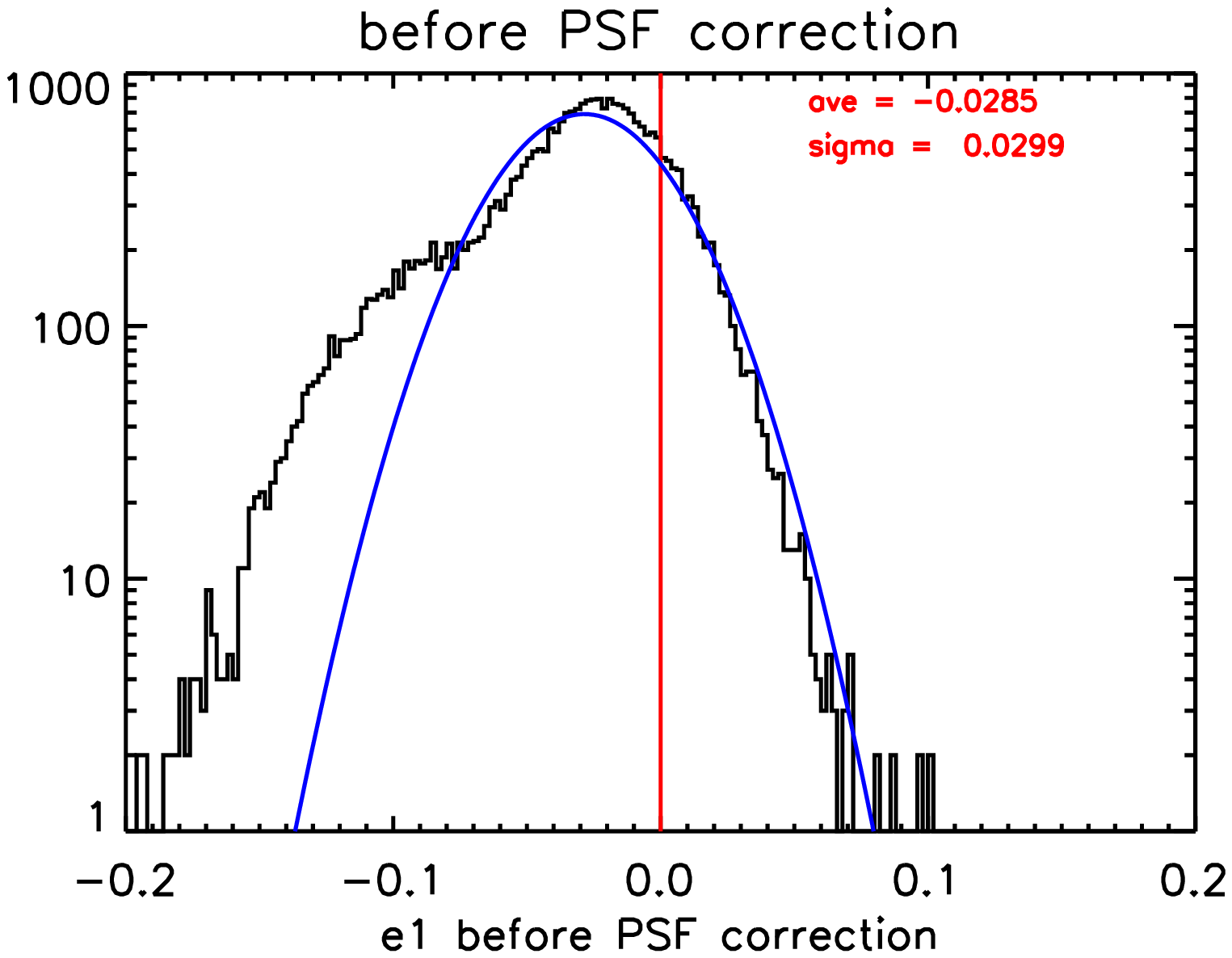}{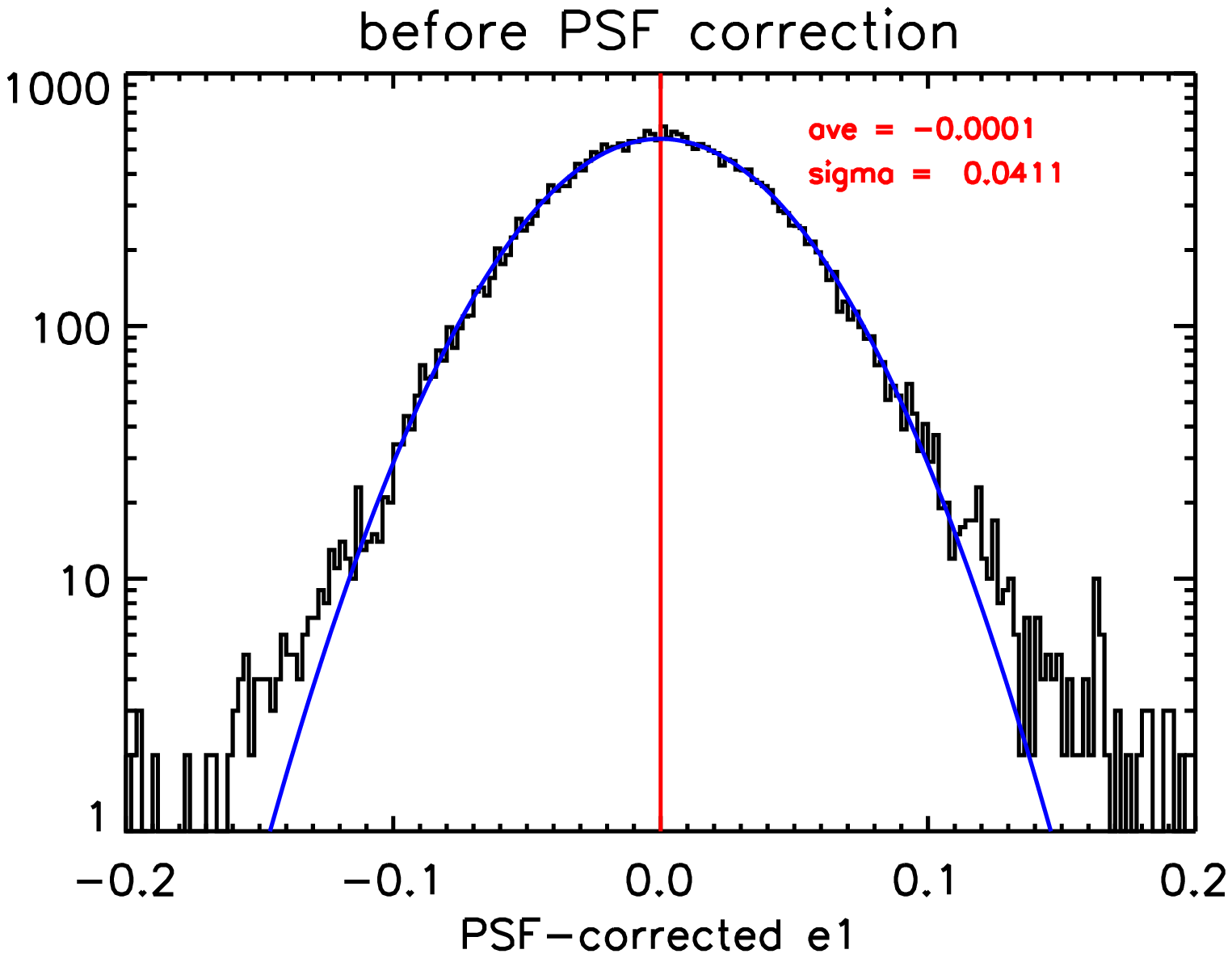}{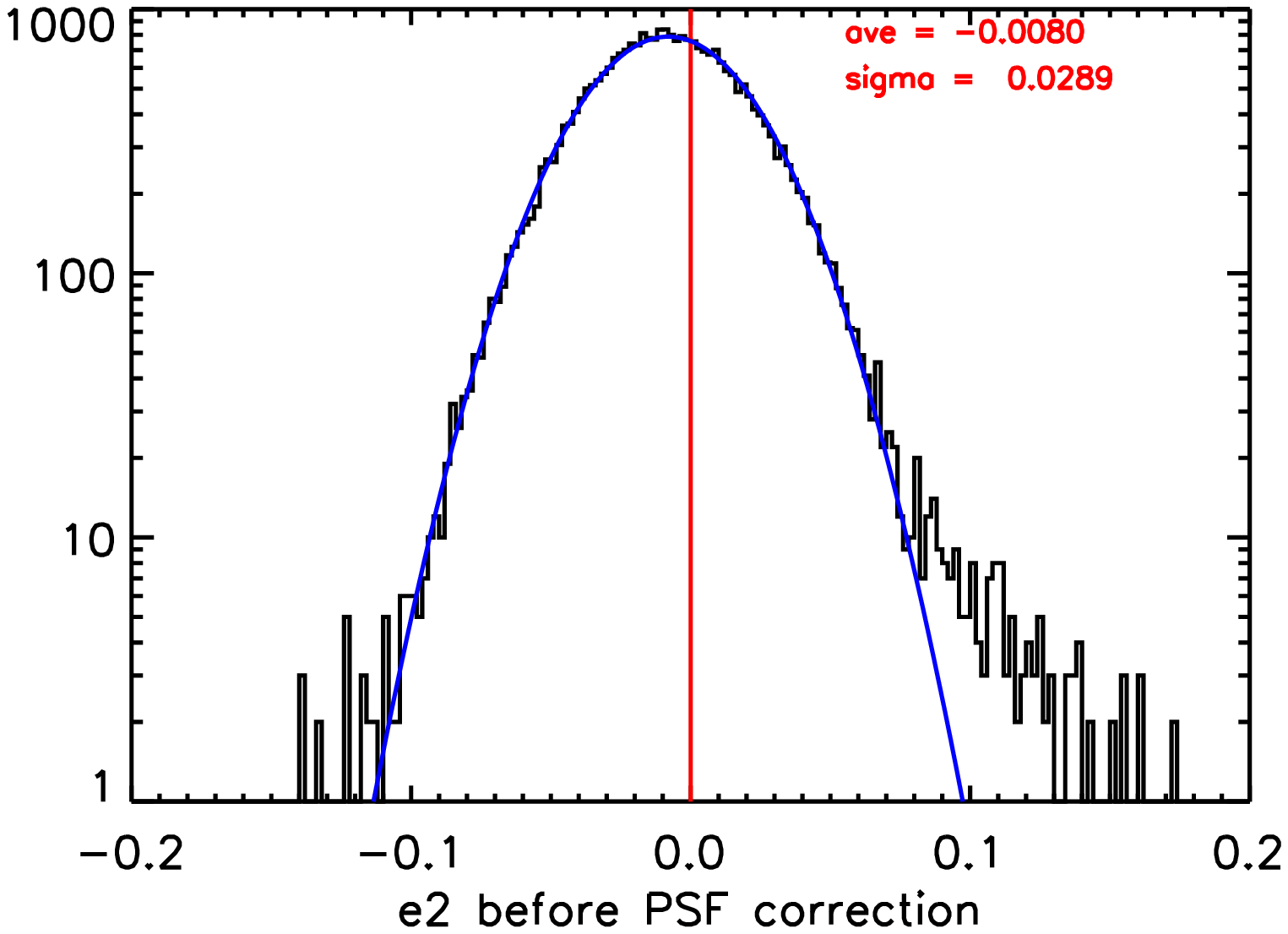}{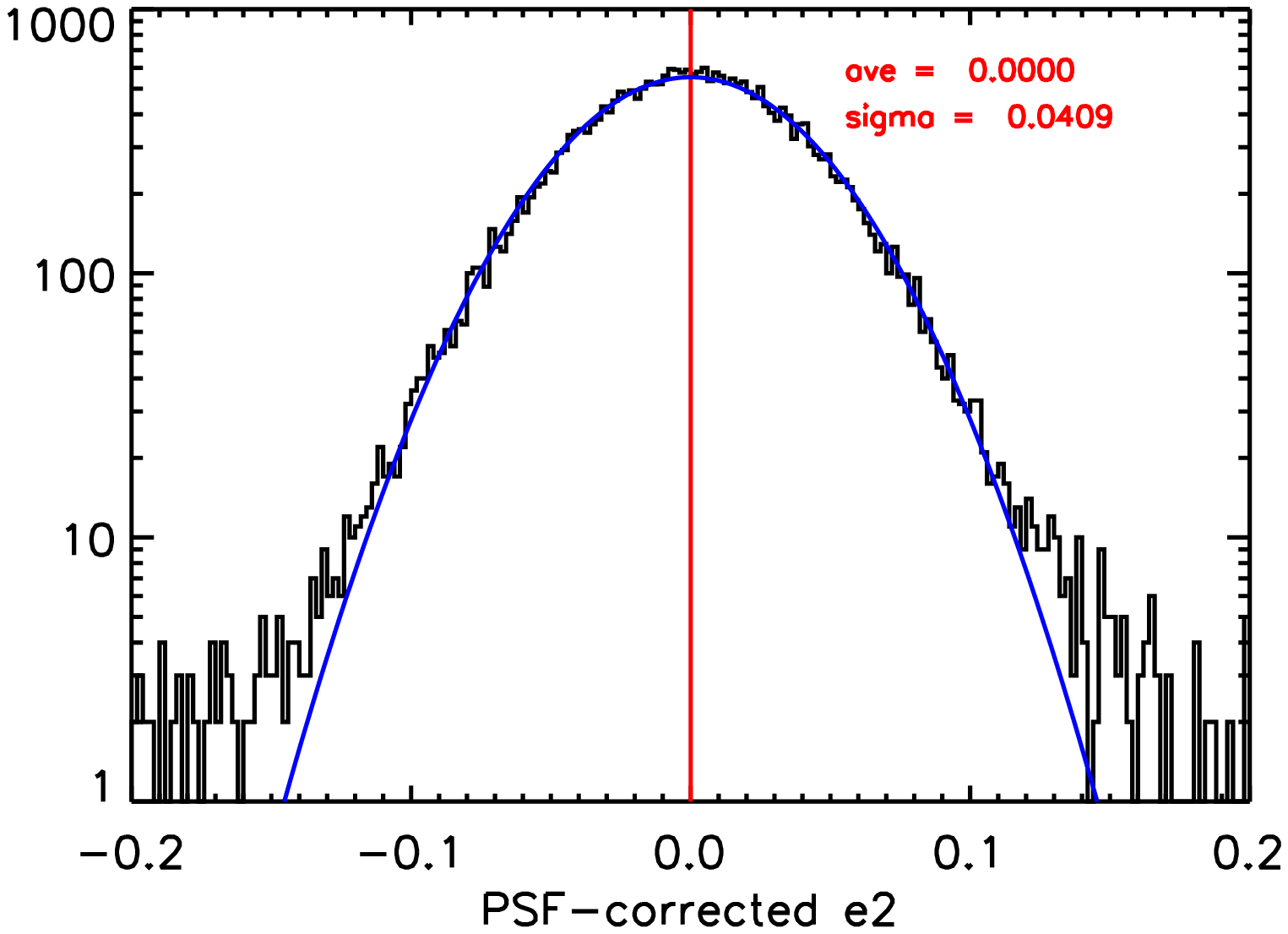}{The
distribution (black) of galaxy ellipticities $e_1$ (top) and $e_2$ (bottom), 
averaged over the $0.1^\circ \times 0.1^\circ$ pixels used in our analysis,
for the photo-$z$ error $< 0.15$ sample.  
The distributions are shown both before (left) and after (right) PSF 
correction, in order to demonstrate the effectiveness of our procedure
in removing the PSF from the galaxy ellipticity measurements.
Also shown are best-fit Gaussians (blue), which are good approximations 
to the PSF-corrected ellipticity distributions.  The text
in red gives the mean and $\sigma$ of the Gaussian in each panel.}

\section{Two-Point Functions}\label{sec:theory}
\newcommand\cross{\times}

The measured ellipticities, $e_1$ and $e_2$, receive contributions (in addition to the noise) from the cosmic shear components $\gamma_1$ and $\gamma_2$.  
The mean cosmic shear is zero, but the correlations are non-zero and dictated by the underlying cosmology. The most basic two-point function is obtained by
multiplying the shears ($e_i/R$) of pairs of galaxies, collecting pairs with angular separations in a given bin to form, e.g., $\xi_{ii}(\theta)$. This two point function depends
on the power spectrum of the convergence field $\kappa$; for example, the expected value of the correlation function for $\gamma_1$ between two pixels located at angular positions $\vec n_1$ and $\vec n_2$
is~\citep{Hu:2000ax}
\bea 
\langle\gam{1}{\vecn{1}}\gam{1}{\vecn{2}}\rangle&=& \int \frac{d^2l}{(2\pi)^2} C_l^{EE} \cos^22\varphi_l \vs &\times&\left[ j_0(l_x\sigma/2)j_0(l_y\sigma/2)\right]^2 e^{i\vec l\cdot (\vec n_1-\vec n_2)}
\eql{g1g1}\eea
where $\varphi_l$ is the angle that the 2D vector $\vec l$ makes with a fixed x-axis and the spherical Bessel functions encode the effects of 
the window function of a square pixel with sides $\sigma$. The power spectrum of the convergence is 
\begin{equation}
C_l^{EE} = \int_0^\infty d\chi \frac{W^2(\chi)}{\chi^2} P_\delta(k=l/\chi;z(\chi)).
\eql{pkappa}
\end{equation}
where $\chi(z)$ is the comoving distance out to redshift $z$, $P_\delta$ is the matter power spectrum, and the window function
depends on the distribution of background galaxies:
\begin{equation}
W(\chi) = \frac{3}{2} \Omega_m H_0^2 \chi \int_{\chi}^\infty d\chi' \frac{dn}{d\chi'} \left( 1-\frac{\chi}{\chi'}\right)
\end{equation}
with $H_0$ the current Hubble rate.
For the redshift distribution of Stripe 82, the window function is a smooth function peaking at $z\simeq0.35$ and is shown in Fig.~\rf{w}.

\Sfig{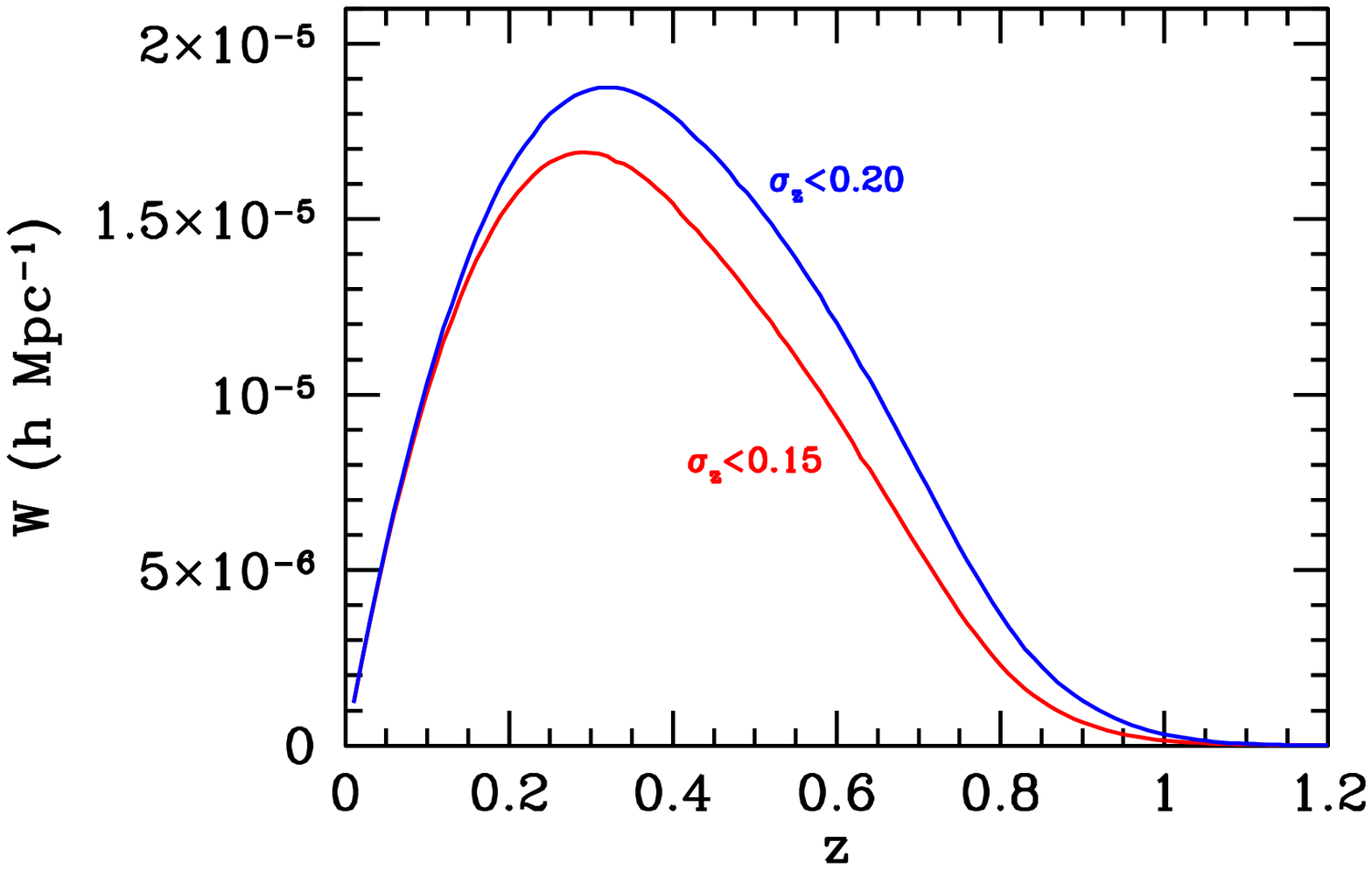}{The window function that weights the power spectrum in \ec{pkappa} for background galaxies in the two cuts used in our sample.}

Note that the 2-point function of $\gamma_1$ in \ec{g1g1} depends not only on the angular distance between two pixels, but also on the direction of this vector $\vec n_1-\vec n_2$. It is therefore useful to combine the various 2-point functions of $\gamma_1$ and $\gamma_2$ into two that depend only on the distance, one of which is not sourced by the convergence and so should vanish, and the other of which contains all the information about the power spectrum of the convergence. Towards this end, we decompose the shears into tangential and cross components measured relative to the line connecting a galaxy pair. The tangential component is perpendicular (positive) or parallel (negative) to this line, while the cross-component has axes 45 and 135 degrees away from this line. The correlation function of the two components are then built by summing the products of $e_te_t$ and $e_\cross e_\cross$ for each pair separated by an angular distance within the bin of interest. 
One can then show that, of the two linear combinations
\be 
\xi_{\pm}(\theta) \equiv \xi_{tt}(\theta) \pm \xi_{\cross\cross}(\theta)\eql{xipm},
\ee
$\xi_{+}$ is simply equal to 
\be
\xi_{+}(\theta) =
\int \frac{d^2l}{(2\pi)^2} C_l^{EE} \left[ j_0(l_x\sigma/2)j_0(l_y\sigma/2)\right]^2 \cos\left[ l_x\theta\right]
.\ee

The goal is to separate modes produced by scalar cosmological perturbations (so-called E-modes) from those produced by systematics (B-modes). 
Several groups have shown~\citep{Schneider:2001af,Fu:2007qq} that the cleanest way to do this is to define
\begin{equation}
\xi_{E,B}(\theta) = \frac{\xi_+(\theta)\pm\xi'(\theta)}{2}.
\eql{xi}\end{equation}
where
\begin{equation}
\xi'(\theta) \equiv \xi_-(\theta) + 4\int_0^\infty \frac{dx}{x} \xi_-(x)
\left(1-3\frac{\theta^2}{x^2}\right).
\end{equation}
For scalar perturbations, $\xi_B=0$, and $\xi_E=\xi_+$.
To check for systematics, we will compute all of these but argue that $\xi_+$ may be slightly polluted by a small B-mode contamination
so we obtain final cosmological constraints from $\xi_E$.

An alternative approach is to extract the $C_l$ spectra directly from a quadratic estimator of the observed shears, essentially inverting \ec{g1g1} and its cousins.
In \S\ref{sec:power}, we describe algorithms for this direct approach of extracting the $E$- and $B$-spectra.

\section{Correlation Function Results}\label{sec:xi}

\subsection{Simulations} 

To set up expectations, we generated mock catalogs of shear for the 
Stripe 82 area, first divided into 42 $2.6^\circ\times2.6^\circ$ boxes, and
then with each box further subdivided into square pixels of size $6'\times 6'$
($ = 0.1^\circ \times 0.1^\circ$).
The cosmic shear in each pixel is drawn from a Gaussian distribution with mean zero and covariance matrix 
that includes both signal and shape noise. The signal part of the covariance matrix accounts 
for all correlations over the larger box and is computed using \ec{g1g1} and similar 2-point functions for $\gamma_2$
assuming a standard $\Lambda$CDM model ($H_0=70$ km s$^{-1}$ Mpc$^{-1}$; $\Omega_m=0.25$, $\sigma_8=0.8$)
and the redshift distribution depicted in Fig.~\rf{dndz}. 
Shape noise adds a diagonal term to the shear covariance matrix equal to 
$(e_{\rm int}/R)^2/N_i$, where the intrinsic shape noise 
$e_{\rm int} = 0.37$ and the shear responsivity $R=1.7$, as noted earlier in
Section~\ref{sec:shape_psf}.  The number of galaxies $N_i$ in the $i$-th 
small pixel is set equal to the number of galaxies in the real data in that
same pixel with one caveat. The average density of galaxies per pixel in the mock catlog is 264, similar to the SDSS data set
with no photo-$z$ cuts. The density in the catalog with $\sigma_z<0.15$ is more than a factor of two smaller than this. So the
errors estimated using the mocks should be smaller than those we eventually obtain using the cut photo-$z$ samples.

We generate 23 full Stripe 82 mock surveys, each of which is divided into
42 $2.6^\circ\times2.6^\circ$ boxes, for a total of 966 boxes.
For each of these boxes we compute the shear-shear correlation function
vs.\ $\theta$, the angular separation between the $0.1^\circ \times 0.1^\circ$
spatial pixels into which the simulated shears are binned.  
We first use the following estimator for the tangential ($\xi_{tt}$) and
cross ($\xi_{\times\times}$) correlation functions
\begin{eqnarray}\eql{xi_tx}
\xi_{tt}(\theta) & = &
\frac{\sum_{i,j}{N_i N_j \gamma_{t,i} \gamma_{t,j}}}{\sum_{i,j}{N_i N_j}} \\
\xi_{\times\times}(\theta) & = &
\frac{\sum_{i,j}{N_i N_j \gamma_{\times,i} \gamma_{\times,j}}}{\Sigma_{i,j}{N_i N_j}} \nonumber
\end{eqnarray}
where the sum is over all pairs $i,j$ of pixels
separated by the angle $\theta$, $\gamma_{t,i}$ and $\gamma_{\times,i}$ are
the tangential and cross components of the shear in pixel $i$
(related to the ellipticity by $\gamma = e/R$),
and $N_i$ is the number of galaxies in pixel $i$.
The weighting by $N_i$ is equivalent to inverse variance weighting according
to the number of galaxies in each pixel, taking each mock galaxy to 
contribute a fixed amount of intrinsic shape noise 
$\sigma_\gamma = e_{\rm int}/R$ per shear component.  
We then compute the $\xi_+$ correlation function using \ec{xipm} 
and plot the results in Fig.~\rf{ximock}, where we show 
the mean (and standard deviation of the mean) of $\xi_+$ averaged over 
all the mock catalogs, as well as the results from a single, typical 
Stripe 82 mock.  Also plotted as a curve is the input correlation function.  
At small $\theta$, the mean value of $\xi_+$ averaged over all the mocks is 
slightly higher than the input $\xi_+$ due to binning.  
A given angular bin contains only a fixed number
of separations because the pixel centers are spaced periodically.  
For example, the lowest bin, shown as centered on $\theta=0.13^\circ$,
contains only the two separations $0.1^\circ$ and $0.14^\circ$.  
A proper treatment would weight these two contribution appropriately and
eliminates the small discrepancy.  
However, this weighting produces only very small changes in $\xi$, 
changes that we are sensitive to only when considering the average over all the 
mocks, but which are far below the noise in the actual data set, as
can be seen in Fig.~\rf{ximock}.

\Sfig{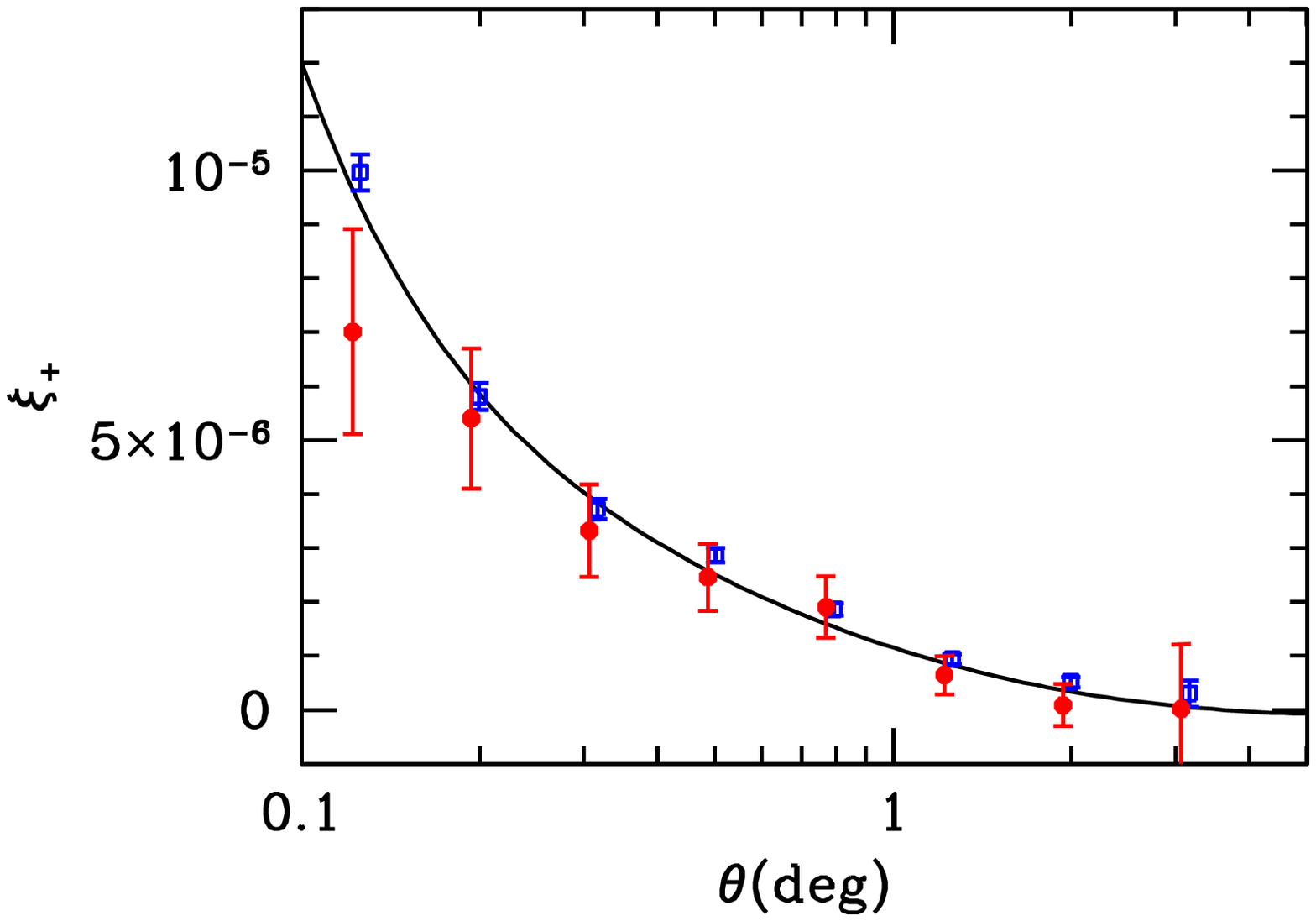}{Correlation function $\xi_+$ measured for 23 mock catalogs, each
consisting of the full Stripe 82 area.  The solid curve is the input 
true correlction function.  The rectangular open (blue) points and error bars
show the mean and standard deviation of the mean of $\xi_+$ averaged over
all 23 mocks, while the circular closed (red) points and errors are the 
results for a single, typical mock.}

We then extract cosmological parameters from each of the 23 
mock surveys.  To do this, we form a $\chi^2$ with the covariance matrix 
measured from the 42 boxes comprising each Stripe 82 mock catalog. 
We then scan over a range of values of $\sigma_8$ and $\Omega_m$, and 
Fig.~\rf{bffinal} shows the resulting best-fit $\sigma_8$ and $\Omega_m$ 
values for all the mocks, along with the true values for the input model. 

\Sfig{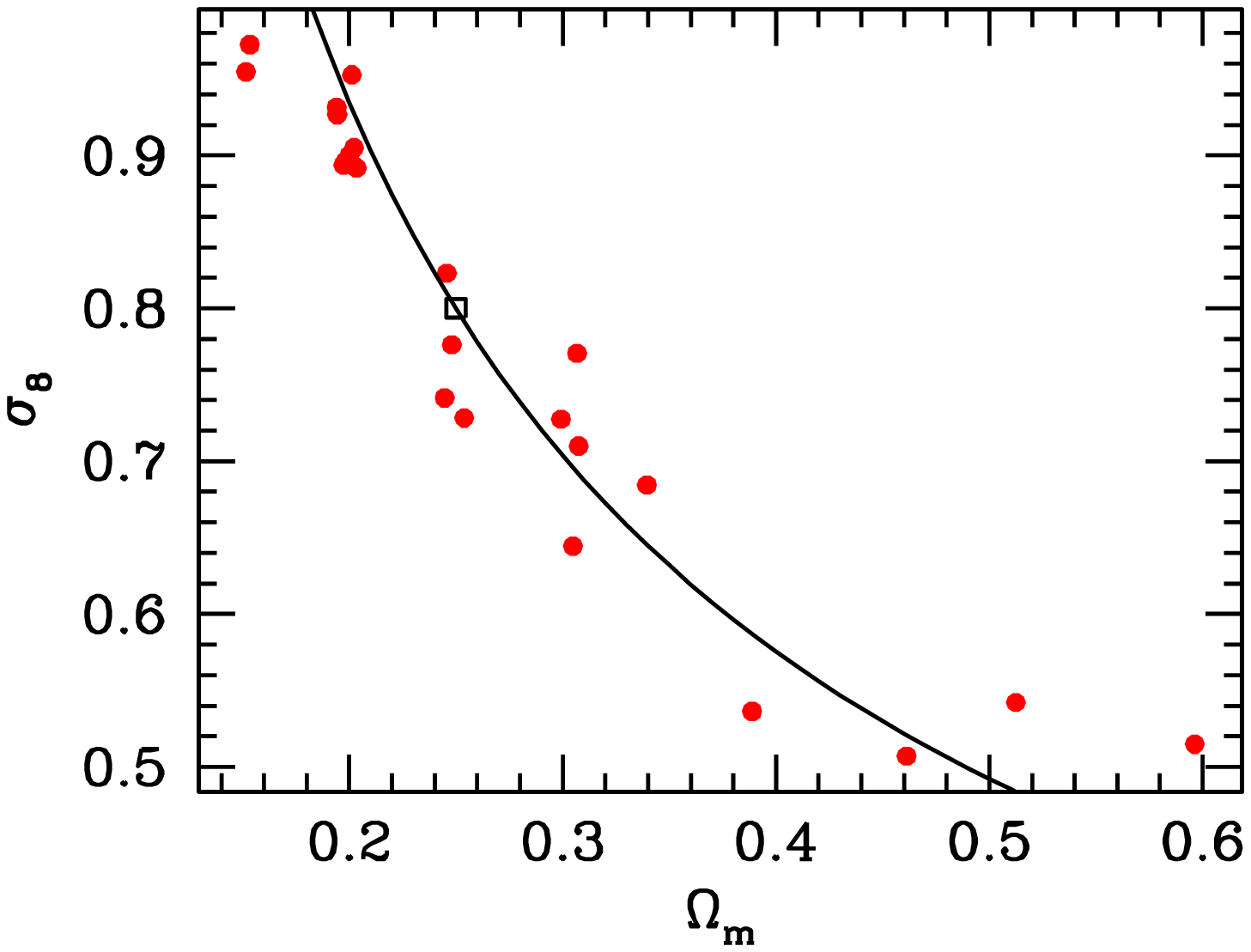}{Best fit values of the cosmological parameters $\sigma_8$ and 
$\Omega_m$ from 23 mock catalogs, each simulating the full Stripe 82 area, 
analyzed with the correlation function.
The open square denotes the true input model. The line traces out 
the degenerate direction, with $\Omega_m^{0.7}\sigma_8$ constant.}

Fig.~\rf{bffinal} suggests a convenient way to quote the results of both the mocks and the data. Apparently only a combination of the two parameters is constrained by the data, with the
degeneracy given by $\Omega_m^{0.7} \sigma_8=$ constant. The input value of this combination (with $\Omega_m=0.25$ and $\sigma_8=0.8$) was $\Omega_m^{0.7}\sigma_8=0.303$, while the mean over all mocks was $0.300$. The rms of the mocks was $0.022$, consistent with the error bar from a single mock. 

\subsection{Data} \label{sec:xi_data}

We now compute the shear-shear
correlation functions for the real data and plot them in
Fig.~\rf{xi_data}.  
We use the same estimators for the tangential ($\xi_{tt}$) and
cross ($\xi_{\times\times}$) correlation functions given earlier in \ec{xi_tx}.
The weighting is again inverse variance weighting according
to the number of galaxies in each pixel, now taking each real galaxy to 
contribute a fixed amount of noise $\sigma_\gamma = \sigma_e/R$ per 
shear component.  
As noted earlier in Section~\ref{sec:shape_psf}, for our galaxies 
$\sigma_e = 0.45$ and includes both intrinsic shape noise and the ellipticity 
measurement error.

We then compute the $\xi_+$ correlation function using \ec{xipm}
and the $E$- and $B$-mode correlation functions via \ec{xi}, and plot
the results for these three correlation functions in Fig.~\rf{xi_data},
for each of our photo-$z$ error $< 0.15$ and $< 0.2$ samples.
As we did for the simulations, for the real data we also compute the 
correlation functions by first dividing the Stripe 82 area into 42
non-overlapping $2.6^\circ \times 2.6^\circ$ square boxes, then calculating
$\xi_+$, $\xi_E$, and $\xi_B$ separately for each box, and finally averaging 
the results over all the boxes.  We use this procedure as it allows us to 
easily derive an empirical estimate of the uncertainties and the covariance 
matrix of the data, using the variances and covariances of the $\xi$ values 
determined over the ensemble of 42 boxes.  The correlation functions are 
computed in eight evenly spaced, logarithmic bins of pixel pair separation 
$\theta$, ranging from $0.13^\circ$ to $3.16^\circ$.

The top panels of  Fig.~\rf{xi_data} show that we have a significant 
cosmic shear signal in our data on scales of about $0.1^\circ$ to 
about $1.3^\circ$.  The bottom panels of Fig.~\rf{xi_data} show that the 
individual $\xi_B(\theta)$ values are mostly consistent with zero within the 
error bars, though overall there does appear to be some small positive 
$B$-mode systematic in the data.  This is consistent with the top panels 
of Fig.~\rf{xi_data}, which show that the $\xi_E$ values are generally
somewhat smaller than the $\xi_+$ values; recall from \ec{xi} that
$\xi_+ = \xi_E + \xi_B$.  For the case of no $B$-mode 
contamination, it should suffice to use the more robustly computed $\xi_+$ 
for cosmology fitting, but for our data we will consider both $\xi_+$ and 
$\xi_E$ in the cosmology fits, in order to check for the impact of potential 
$B$-mode contamination.  The best-fit results are also shown in 
Fig.~\rf{xi_data} and will be discussed below in Section~\ref{sec:cosmo}.

\SfigT{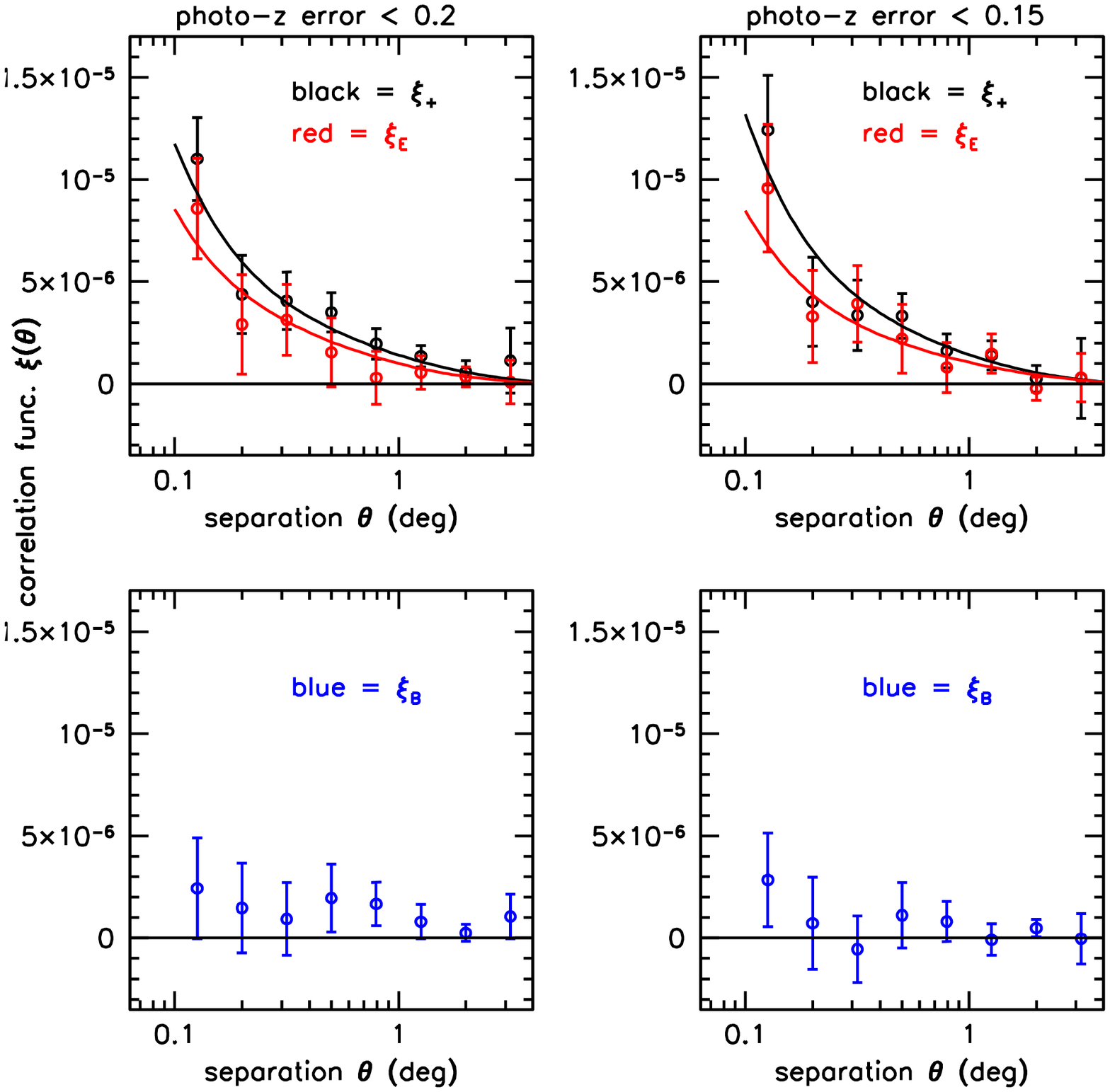}{The measured shear-shear correlation functions 
for our data, plotted as data points with error bars, in logarithmically-spaced
bins of angular separation $\theta$.  The results for $\xi_+$ (black) and 
$\xi_E$ (red) are plotted in the top panels, and the results for
$\xi_B$ (blue) are shown in the bottom panels.  Note the $y$-axis range
is the same in the top and bottom panels to facilitate comparisons.
The left-hand panels are for the photo-$z$ error $\sigma_z < 0.2$ sample
and the right-hand panels for the $\sigma_z < 0.15$ sample.
The curves in the top panels are the correlation functions derived from the 
best-fitting cosmologies to each photo-$z$ error sample, using either the 
$\xi_+$ or $\xi_E$ data.}

\subsection{Systematics Check}\label{sec:systematics}
\begin{figure*}[!htb]
\centering
\includegraphics[width=0.95\columnwidth]{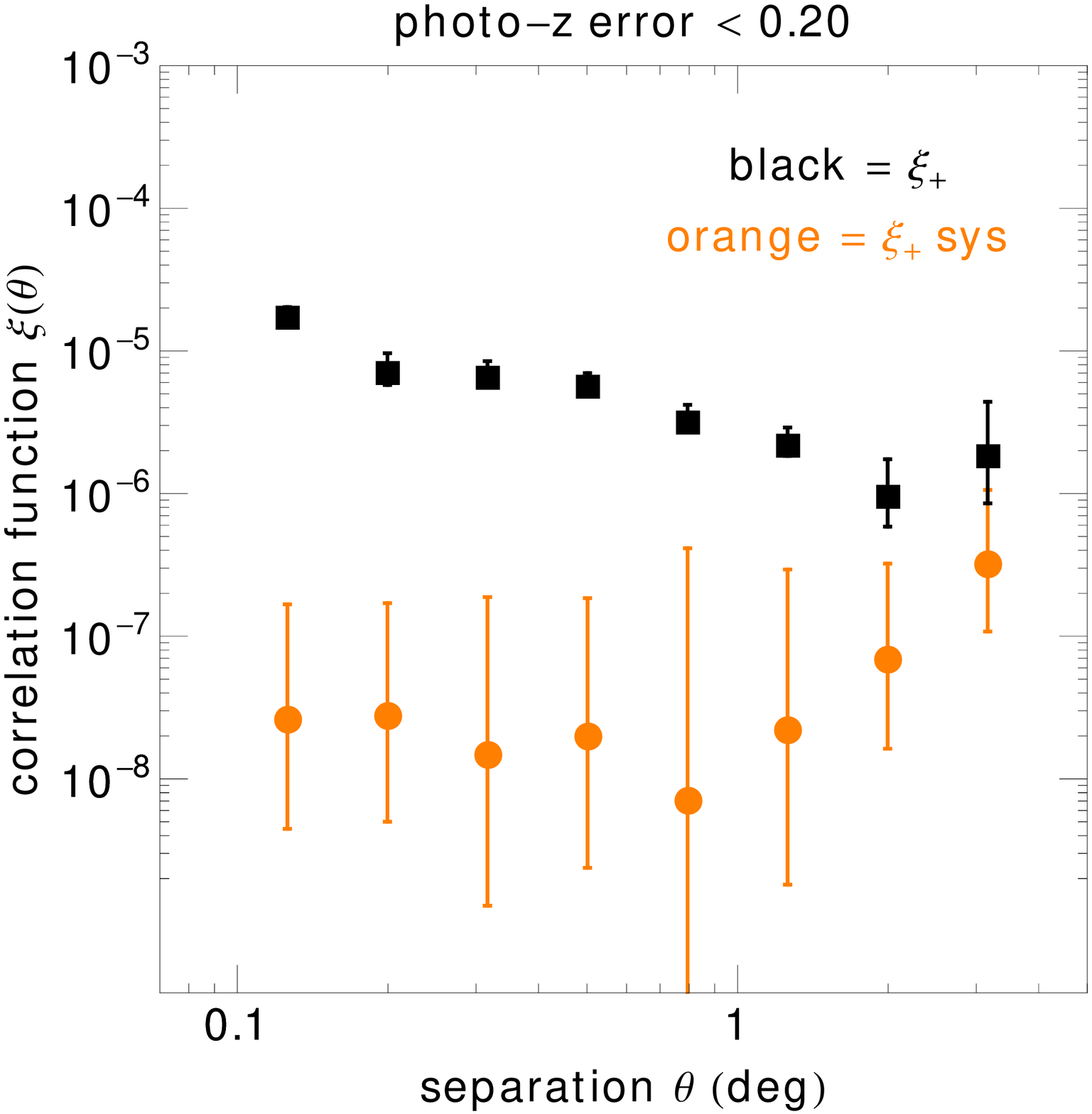}\hspace{0.1\columnwidth}
\includegraphics[width=0.95\columnwidth]{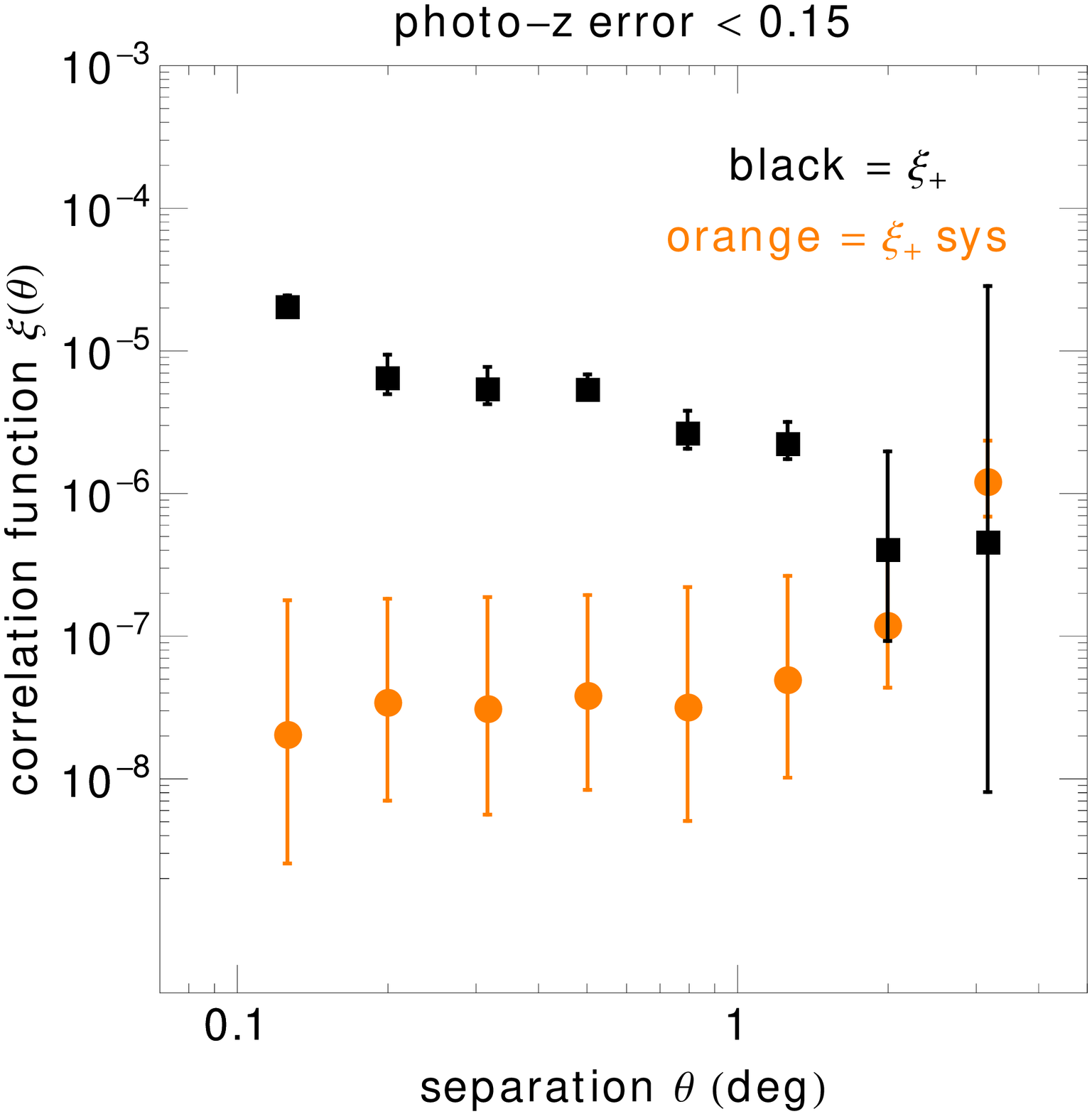}
\caption{Comparison between the correlation function $\xi_{+}^{\mathrm{SYS}}$
induced by residual PSF systematics (orange) and the shear-shear
correlation function $\xi_+$ (black), for both the $\sigma_z<0.20$ (left) 
and $\sigma_z<0.15$ (right) samples, demonstrating that our PSF systematics 
are at a 1\% level on scales $\theta < 2^\circ$.  
See Section~\ref{sec:systematics} for additional discussion.} 
\label{xisys20}
\end{figure*}

Imperfect PSF corrections are the main source of systematics for cosmic shear measurements. 
We can verify that our systematics are under control by computing the 
cross-correlation between the corrected galaxy ellipticities $e$ and the 
uncorrected stellar ellipticities $e^{\star}$, normalized by the 
auto-correlation of $e^{\star}$, following the procedure of 
\citet{Bacon:2003}.  Here we do something similar, but use the PSF model 
ellipticities $e_{\mathrm{PSF}}$ evaluated at the location of each galaxy,
instead of the measured ellipticities of a separate sample of stars.  
This serves the same purpose, i.e., to identify in the measured shear signal $\gamma$
any spurious contributions due to uncorrected contamination from the PSF 
ellipticity. If the observed shear is contaminated by the PSF, then
\begin{equation}
\gamma = \gamma_\mathrm{true} + a \gamma_\mathrm{PSF} \ , 
\end{equation}
where $a$ is a constant. 
Then the observed two-point correlation function gets an unwanted contribution of 
\be 
\xi^{\mathrm{SYS}}_+ =a^2 \langle \gamma_\mathrm{PSF}\gamma_\mathrm{PSF} \rangle
.\ee
The constant $a$ can be estimated by cross-correlating the PSF model ellipticities with 
the galaxy ellipticies leading to \citep{Bacon:2003} 
\begin{equation}
\xi^{\mathrm{SYS}}_{+} = \frac{\langle\gamma \ \gamma_\mathrm{PSF}\rangle^{2}}{\langle
\gamma_\mathrm{PSF} \gamma_\mathrm{PSF}\rangle} \ , 
\end{equation}
where the angular correlation function $\xi_{+}^{\mathrm{SYS}}$ can be 
directly compared to $\xi_{+}$, providing an estimate of the PSF systematic 
effect on our measurement.  
Error bars are computed for $\xi_{+}^{\mathrm{SYS}}$
the same way as for $\xi_+$, using the standard deviation of the mean values
of $\xi$ for the 42 boxes into which Stripe 82 is divided.
Figure  \ref{xisys20} shows the result of 
this check for both the $\sigma_z<0.15$ and $\sigma_z<0.2$ samples, 
demonstrating that our PSF systematics are at about the 1\% level on scales 
$\theta < 2^\circ$.  
For the two largest $\theta$ bins, $\xi_{+}^{\mathrm{SYS}}$ appears to be 
larger in proportion to $\xi_+$, but are still small compared to the 
larger $\xi_+$ errors on those scales.  We have explicitly compared
the cosmology fits with and without first subtracting off 
$\xi_{+}^{\mathrm{SYS}}$ from $\xi_+$ and have found negligible differences,
thus confirming that our cosmology results are not significantly
contaminated by residual PSF systematic errors.

In addition to PSF-related systematics, uncertainties in the redshift
distribution of our galaxy sample will also lead to errors in our cosmic
shear results.  As discussed earlier in Section~\ref{sec:photoz}, we address
this issue by presenting our results using two galaxy samples defined by
the different photo-$z$ error cuts $\sigma_z < 0.2$ and $\sigma_z < 0.15$.
As will be shown in Section~\ref{sec:cosmo}, our cosmology results are
insensitive to the choice of photo-$z$ cut.

Also, as described in Section~\ref{sec:shape_psf}, we found and corrected
for a residual additive bias in the shear measurement, which is likely due
to CCD camcol-dependent errors in the PSF model affecting the subsequent 
linear PSF correction and galaxy ellipticity measurements.  
In addition to this additive bias, the \citet{Hirata:2003} linear PSF 
correction scheme we use is subject to a multiplicative shear calibration error 
$\delta\gamma/\gamma$ that depends on galaxy type and resolution factor.
This issue was studied in detail for SDSS galaxies in \citet{Hirata:2003} and 
\citet{Hirata:2004}, with the latter concluding that 
$|\delta\gamma/\gamma| \lesssim 0.07$.  This fractional shear calibration error
leads to a fractional error in the correlation function
$|\delta\xi / \xi | \approx 2 |\delta\gamma/\gamma| \lesssim 0.14$.
Compared to the statistical fractional error in the amplitude of $\xi_+$
of about 0.25 (determined as usual from our 42 boxes on Stripe 82),
the additional systematic fractional uncertainty $|\delta\xi / \xi |$
due to shear calibration leads only to a $\lesssim 15\%$ increase in 
the fractional error on $\xi_+$ when added in quadrature.  We will neglect 
this as it is a small effect compared to the much larger existing statistical 
errors on $\xi$.

\section{Power Spectrum}\label{sec:power}

\def \ecut [#1]{|e_1|,|e_2|< #1}
\def\rmse{\sigma^2_{e_{1,2}}}
\def\LCDM{\Lambda CDM}
\def\rmstl{\sigma^2_{\gamma}}
\def\rmst{\gamma^2}
\def\Ngal{N_{\rm gal}}
\def\om{\Omega_m}
\def\Cov{\mathbb{C}}
\def\Win{\mathbb{W}}
\def\pal{\mathcal{D}_{\alpha}}
\def\pan{\mathcal{D}_{N}}
\def\llan{\left\langle}
\def\rran{\right\rangle}
\def\dzs{\sigma_z < 0.15}
\def\dzl{\sigma_z < 0.2}
\def\scr{\textcolor{red}}
The power spectrum and the correlation function are complementary ways of measuring the two point statistics of any field.
The power spectrum method, however, has been used less commonly for cosmic shear \citep[but see][]{Brown:2002wt},
due to the complexity in accounting for the non-trivial window function in Fourier space that 
arises from survey geometry and from a non-uniform galaxy distribution.
 In this section we 
estimate the power spectrum of the Stripe 82 coadded data, while accounting 
for the effects of the survey geometry and of spatially non-uniform shape noise. 
We use the quadratic estimator (hereafter ``QE'') to estimate the power spectrum and its errors, following the approach proposed by 
\cite{Hu:2000ax} \citep[also see][]{Seljak:1998wl}.
To cross-check our answer, we also derive the power spectrum using a fast Fourier transform,
taking advantage of the nearly flat geometry of Stripe 82, and construct a pseudo estimator (hereafter ``PE'') of the power spectrum.
For the pseudo estimator, we first weight the shear 
value measured in each pixel with the inverse shape noise in order to 
down-weight masked pixels or pixels with large shape noise, then Fourier transform, 
generate a pseudo power spectrum, and 
finally deconvolve the effect of the weight to extract an unbiased power 
spectrum. 
We test the fidelity of these methods
using the 23 Gaussian SDSS mock catalogs that mimic the Stripe 82 data,
 and then apply 
the two methods 
to the 
observed real Stripe 82 data. In order to ease the computational 
load in calculating matrix operations, we divide the Stripe 82 data into 42 boxes
of $2.6^\circ \times 2.6^\circ$ and ignore the correlations between different boxes. 
That is, we lose clustering information on scales larger 
than $\sim 1.3^\circ$.

\subsection{Quadratic Estimator}
\subsubsection{Summary of Method}
We follow the method described in \cite{Hu:2000ax}, which is summarized below.  
We first assume that the likelihood of the measured shear field is:
\begin{eqnarray}
\mathcal{L}&=&\frac{1}{(2\pi)^N|\Cov(\pal)|^{1/2}}\exp\left[-\frac{1}{2}d^T\Cov^{-1}(\pal)d\right], 
\end{eqnarray}
where $d$ is the data vector, i.e., the two components 
of the measured shear field for the 42 boxes 
of $26\times 26$ pixels.
The covariance matrix $\Cov$ is the sum of the cosmological 
signal $\Cov^{\rm sig}$ and the noise $\Cov^{\rm noise}$ due 
to the rms intrinsic ellipticity and the measurement errors:
\begin{eqnarray}
\Cov&=&\Cov^{\rm sig}+\pan \Cov^{\rm noise},
\end{eqnarray}
where we have inserted a parameter $\pan$ to account for any deviation from the assumed level of shape noise; that is, if the
shape noise is known perfectly, then $\pan$ is fixed to one. 
 In the case of data with large shape noise, we find that $\pan$ and the EE and BB power from the cosmic signal 
 are difficult to estimate simultaneously, so we will fix $\pan$ to unity. 
But it is useful to allow it to vary in the case 
 of relatively low shape noise.
Since we are ignoring the correlation between the boxes,
 $\Cov$ is block-diagonal, with 42 separate blocks. The signal part is identical from box to box, but
 because the galaxy density varies, the shape noise
 differs slightly from one block to another.

\begin{figure*}[thbp]
\sfig{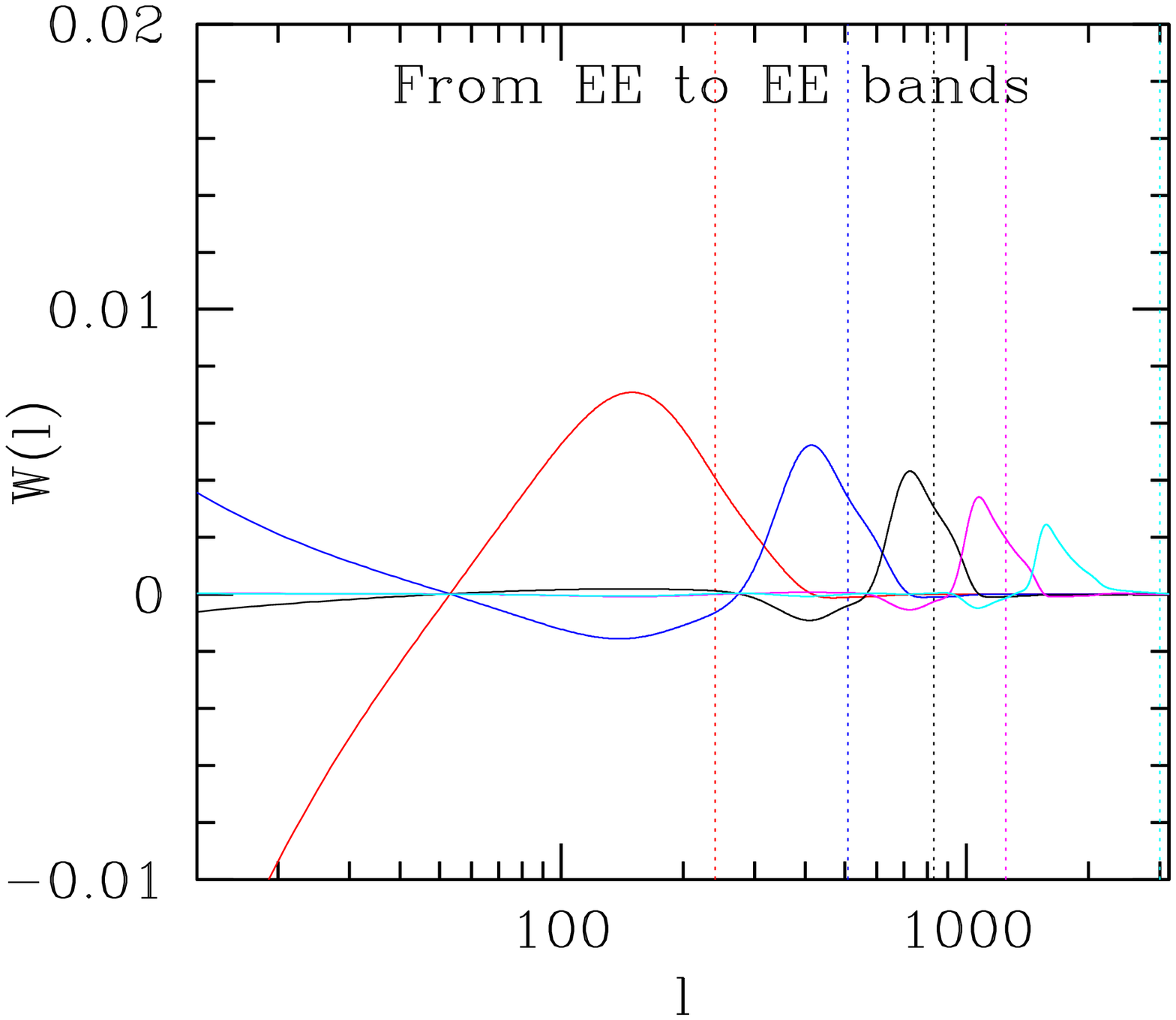}{1.05\columnwidth}
\sfig{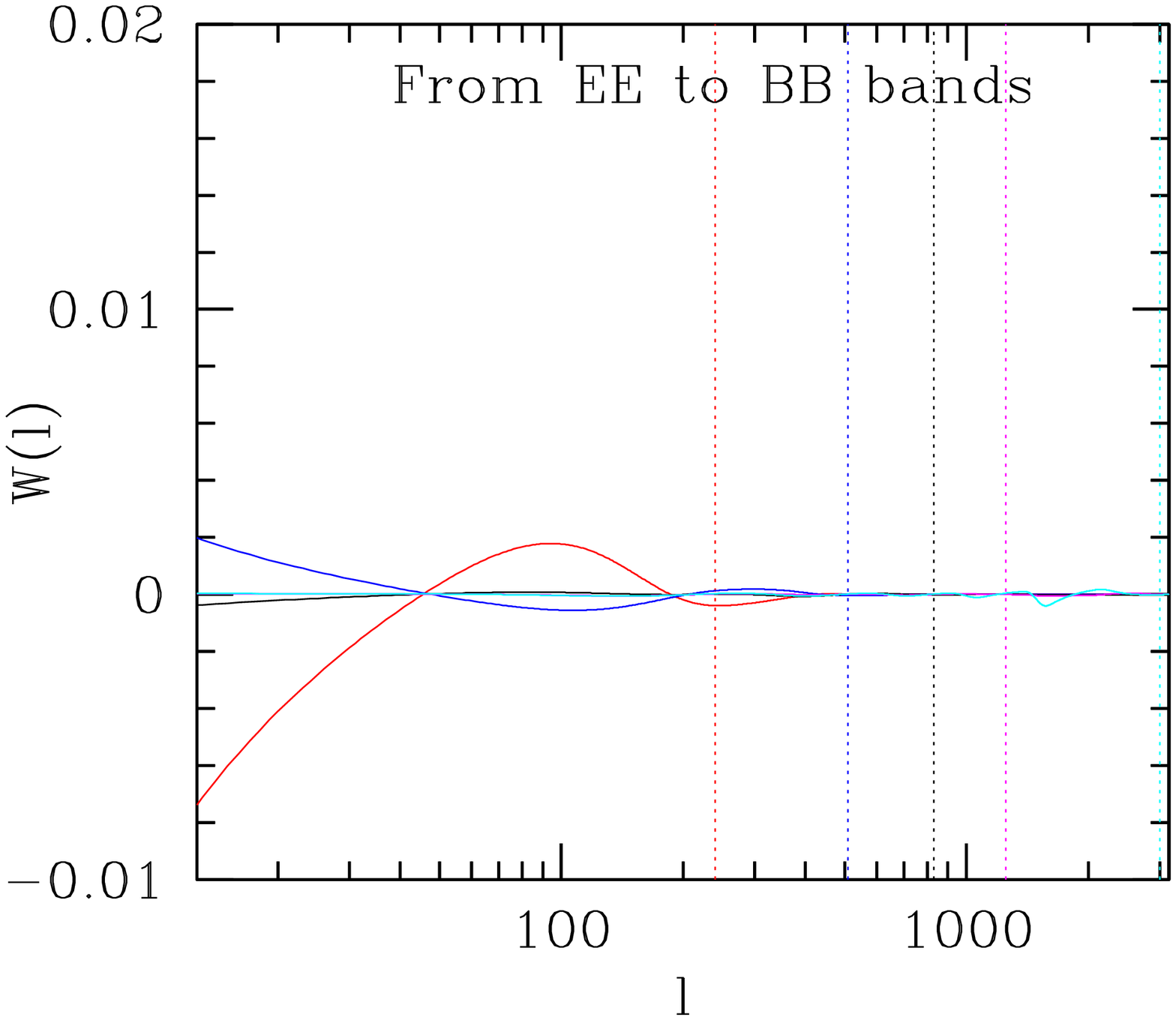}{1.05\columnwidth}
\caption{Band window functions. The left panel shows the contribution from the EE power spectrum at an integer wave 
number to the EE band power. The curves with different colors describe different bands and the vertical lines 
correspond to the naive centers of the bands. The sum of each curve is close to unity. The right panel shows 
the contribution from the EE power spectrum to the BB band power. As expected, there is no leakage between EE 
and BB modes on average.}
\label{fig:window}
\end{figure*}

Labeling the shear components with indices $a,b$ and the pixels with $i,j$, the covariance matrices are 
\begin{eqnarray}
&&\Cov^{\rm sig}_{(ij)(ab)}=\langle\gam{a}{\vecn{i}}\gam{b}{\vecn{j}}\rangle,\nonumber\\
&&\Cov^{\rm noise}_{(ij)(ab)}=\frac{\sigma_\gamma^2}{N_i}\delta_{ij}\delta_{ab}, 
\end{eqnarray}
where $N_i$ is the number of galaxies in pixel $i$ and $\sigma_\gamma$ is the rms of the shears of all the galaxies.

Eq. \Ec{g1g1} gives the expression for one element of the signal covariance matrix under the assumption that
only the $E$-mode is non-zero. We want to simultaneously measure the $E$- and $B$- modes (using the latter as 
a systematics check), so we need to generalize \ec{g1g1}. Following \cite{Hu:2000ax}, 
we write
\begin{eqnarray}
&&\langle\gam{1}{\vecn{i}}\gam{1}{\vecn{j}}\rangle = \int \frac{d^2l}{(2\pi)^2} [\Cee(l)\cos^22\varphi_l \nonumber \\
&&+\Cbb(l)\sin^22\varphi_l-\Ceb(l)\sin 4\varphi_l]W^2(\vec{l})e^{i\vec{l}\cdot(\vecn{i}-\vecn{j})},\nonumber \\
&&\langle\gam{2}{\vecn{i}}\gam{2}{\vecn{j}}\rangle = \int \frac{d^2l}{(2\pi)^2} [\Cee(l)\sin^22\varphi_l \nonumber \\
&&+\Cbb(l)\cos^22\varphi_l+\Ceb(l)\sin 4\varphi_l]W^2(\vec{l})e^{i\vec{l}\cdot(\vecn{i}-\vecn{j})}, \nonumber \\
&&\langle\gam{1}{\vecn{i}}\gam{2}{\vecn{j}}\rangle = \int \frac{d^2l}{(2\pi)^2} [ \frac{1}{2}(\Cee(l)-\Cbb(l))\sin 4\varphi_l \nonumber \\
&&+\Ceb(l)\cos 4\varphi_l ] W^2(\vec{l}) e^{i\vec{l}\cdot(\vecn{i}-\vecn{j})},
\end{eqnarray}
where $\varphi_l$ is the angle between $\vec{l}$ and the $x$-axis, $W(\vec{l}) = j_0(l_x\sigma/2)j_0(l_y\sigma/2)$ is the pixel window function 
in Fourier space, and $\sigma$ is the pixel side ($0.1^\circ$) in radians.

We now approximate the angular power spectra with piecewise constant band powers; that is, we set $l(l+1)C(l)/2\pi$ to a
constant value $\pal$ over a band $\alpha$ spanning a range of $l$. Then the signal covariance matrix 
is a linear combination of the band powers:
\begin{eqnarray}
\Cov^{\rm sig}_{(ij)(ab)}&=&
\sum_{\alpha} \pal \int_{l\in\alpha} \frac{dl}{2(l+1)} \vs
&&\times \left[ w_0(l) I^{\alpha}_{(ij),(ab)} + \frac{1}{2} w_4(l) Q^{\alpha}_{(ij),(ab)} \right],\label{eq:Covsig}
\end{eqnarray}
where $\pal$ are the $EE$, $BB$, and $EB$ band powers.
The integration in Eq. \Ec{Covsig} runs over the range of $l$ within each band.
We refer readers to \cite{Hu:2000ax} for the exact forms of $w_0$ and $w_4$, which are the 
decomposed pixel window functions, and matrices $I^{\alpha}_{(ij),(ab)}$ and $Q^{\alpha}_{(ij),(ab)}$ 
\footnote{\cite{Hu:2000ax} has a typo in Eq. (14): $Q^{\beta\beta}$ for $\langle\gamma_2\gamma_2\rangle$ is $J_0 + 
2 c_4J_4+ c_8J_8$, not $J_0 + 2 c_4J_4- c_8J_8$.  }.  

 By maximizing the likelihood as a function of the angular power spectrum, i.e., by finding 
 $\Cov^{\rm sig}$ that describes the observed data the best, we derive the best-fit shear angular band powers. 
 The solution is derived iteratively by using the Newton-Raphson method to find the root 
 of $d\mathcal{L}/d\pal=0$, and each step toward an improved estimate is determined by stepping
\begin{eqnarray}
\delta \pal &\propto& \sum_\beta \frac{1}{2}(F^{-1})_{\alpha\beta}{\rm tr}[(dd^T-\Cov)(\Cov^{-1}\Cov_{,\beta} \Cov^{-1}],
\end{eqnarray} 
where $\Cov_{,\alpha}\equiv \partial \Cov/\partial\pal$
and the Fisher matrix is 
\begin{eqnarray}
F_{\alpha\beta}&=&\frac{1}{2}{\rm tr}(\Cov^{-1}\Cov_{,\alpha} \Cov^{-1}\Cov_{,\beta}).
\end{eqnarray}
Assuming that the likelihood is sufficiently Gaussian near the maximum, we interpret $F^{-1}$ as 
the covariance matrix of the measured band powers.

Given an estimated band power, how do we correct for the finite band width and compare it to the theoretical prediction? The simplest ideas -- taking the
value of $\pal$ at the center of the bin or averaging over all $l$'s in the bin -- are not quite right. Rather, 
each measured $\pal$ samples the $C_l$'s with a window function of its own 
\cite[see, e.g.,][]{Knox:1999fg} not exactly
equal to a square well. To compute this, we use the fact that 
the expected value of the band power, $\llan \pal \rran$, is related to the power spectrum at each wave 
number $\mathcal{D}(\ell')$ through the window function $\Win_{\alpha \ell}$:
\begin{eqnarray}
\langle p_{\alpha} \rangle &=& 
\sum_l \Win_{\alpha \ell} \mathcal{D_\ell},
\end{eqnarray}
where $\ell$ is an integer wavenumber (taken here to lie within the range $10<l<3600$), and 
\begin{eqnarray}
\Win_{\alpha \ell}=F_{\alpha \beta}^{-1} \frac{1}{2}{\rm tr}[\Cov^{-1}\Cov_{,\beta}\Cov^{-1}\Cov_{,\ell}], \label{eq:Win}.
\end{eqnarray}
Here $\Cov_{,\ell}$ is the derivative of $\Cov$ with respect to the power at an integer wavenumber $\ell$ 
and derived using Eq. \Ec{Covsig} with $\Delta l = 1$, i.e., without integration. 
When we derive the cosmology fit, we convolve this window function with the theoretical model to compare to the data. 
Fig.~\rf{window} shows the window function for Stripe 82 for our bands. The right panel shows that indeed for the optimal quadratic estimator, our basic systematics
test -- absence of a $B$-mode -- is robust, in that a model with only $E$-modes will {\it not} produce on average a spurious detection of
$B$-modes due to the complicated geometry and non-uniform shape noise.

\begin{figure}[thbp]
\sfig{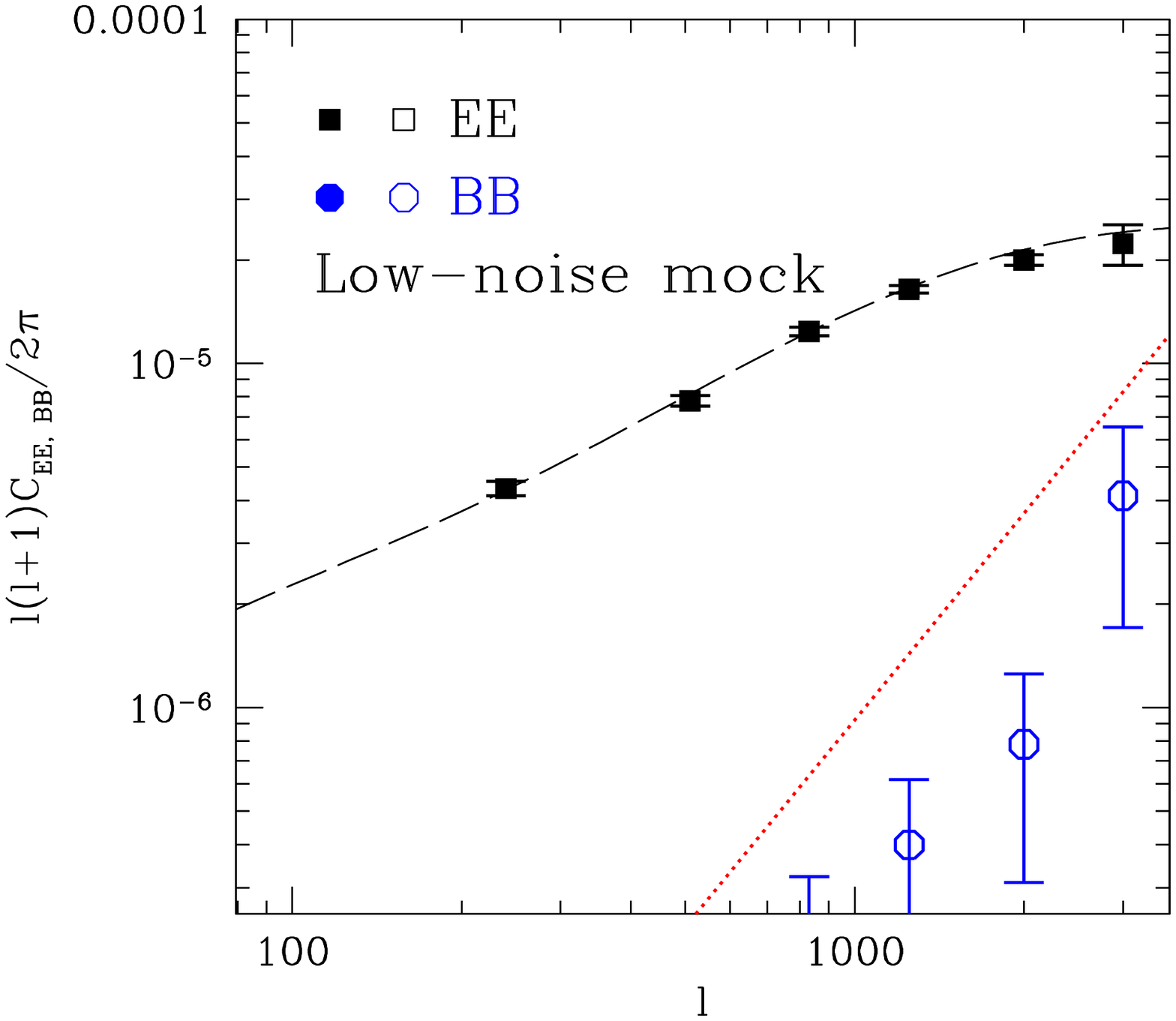}{1.0\columnwidth}
\caption{The best-fit EE (squares) and BB (circle) power spectra of the low-noise mock using the quadratic estimator without prior knowledge of the shape noise. A solid point denotes a postitive value and an open point a negative value. The dashed line corresponds to the input cosmology. The error bars reflect the Gaussian error.  The BB mode is much smaller than the EE mode and is zero within $1-2\sigma$. The dotted line shows the level of shape noise in the power spectrum.}\label{fig:lnqe}
\end{figure}
\begin{figure}[thbp]
\sfig{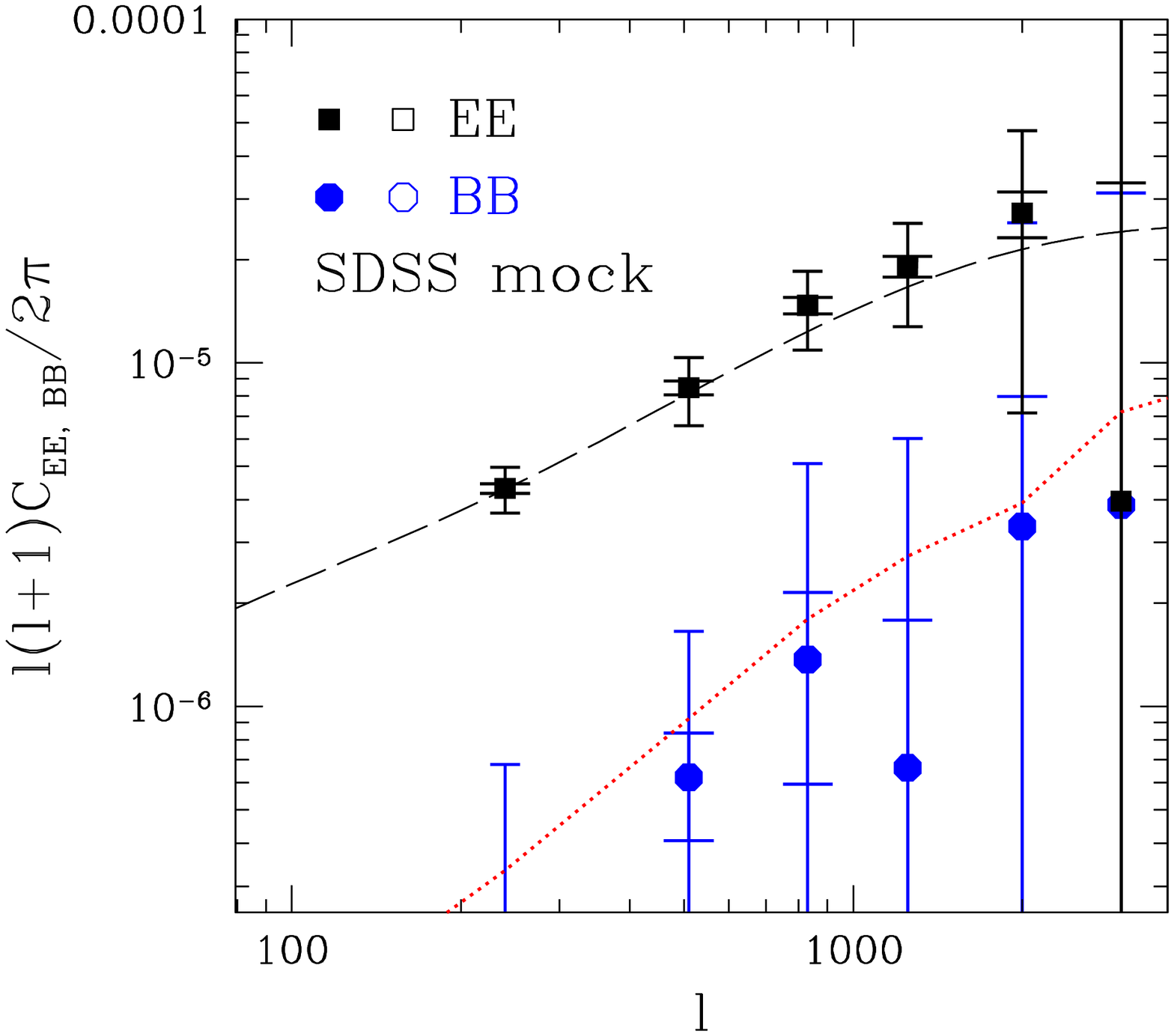}{1.0\columnwidth}
\caption{{\small Power spectra of the SDSS mocks using the quadratic estimator. Left: the average of the best-fit EE and BB power spectra of the 23 SDSS mocks. There are two sets of error bars. The larger set of error bars show the standard deviation among the 23 mocks. The smaller set of error bars show the error associated with the average of the 23 mocks, i.e., the standard deviation divided by $\sqrt{23}$. The dashed line shows the input EE power spectrum. The input BB power is zero. The red dotted line shows the shape noise contribution to the error on the band power. 
}}\label{fig:sdssqe}
\end{figure}

\subsubsection{Mock Tests}
We test our estimator using Gaussian mocks as described in \S\ref{sec:xi}. We analyze two sets of mock catalogs: 
one with very low 
shape noise and one with the pixel-to-pixel varying shape noise that is very similar to the 
observed data on Stripe 82. Hereafter we refer to the former as ``low-noise mocks'' 
and the latter as ``SDSS mocks.''
The low-noise mock is generated using a theoretical Gaussian covariance matrix for the fiducial concordance cosmology, with a small amount
of shape noise, i.e., $\rmstl/\Ngal$, with a constant $\Ngal=250$ and $\rmstl=0.02176^2$; 
we generate one realization of the 42 boxes on Stripe 82.

We first use the low-noise mock to test our estimators in the case of minimal shape noise.  
Figure \ref{fig:lnqe} shows the estimated band power (data points) and the associated 
errors for the EE and BB modes using the low-noise mock. The dashed line corresponds to the input cosmology. 
We find that the EE mode band power is consistent with the input power spectrum to an impressive accuracy. 
The derived best fit BB power spectra are zero within $1-2\sigma$, i.e., fairly consistent with the input.

We next test our methods in the presence of realistic shape noise $\rmstl=0.25^2$, 
using  $\Ngal(\vec{x})$ similar to the real data, as described in \S\ref{sec:xi}.
On average, these SDSS mocks contain $\Ngal=264$ per pixel. We have a total of 23 mocks of the 
42 boxes.

Figure \ref{fig:sdssqe} shows the average and standard deviation of 23 SDSS mock power spectra derived using 
the quadratic estimator. 
We put two sets of error bars on each of the EE and BB band powers. The larger set of error bars corresponds 
to the error associated with one SDSS mock (i.e., the dispersion among the mocks), and 
the smaller set to the standard deviation divided by $\sqrt{23}$, i.e., the error associated with the average.  
Based on the larger set of error bars, one sees that we expect to recover the input power spectrum within 
1$\sigma$ for a survey comparable to Stripe 82. The dispersion among the mocks is 
consistent with the Gaussian error based on the inverse Fisher matrix to within $18\%$.

Fig.~\rf{bfcl} shows the best-fit values of the cosmological parameters extracted from each mock. As in Fig.~\rf{bffinal},
the range of values is consistent with the input model, 
with an average $\Omega_m^{0.7}\sigma_8 = 0.303$, 
the same as the true input value. 
The rms value over all the mocks is 0.022 and is consistent with the 1$\sigma$ error bars
assigned to the parameter in a typical mock.

\Sfig{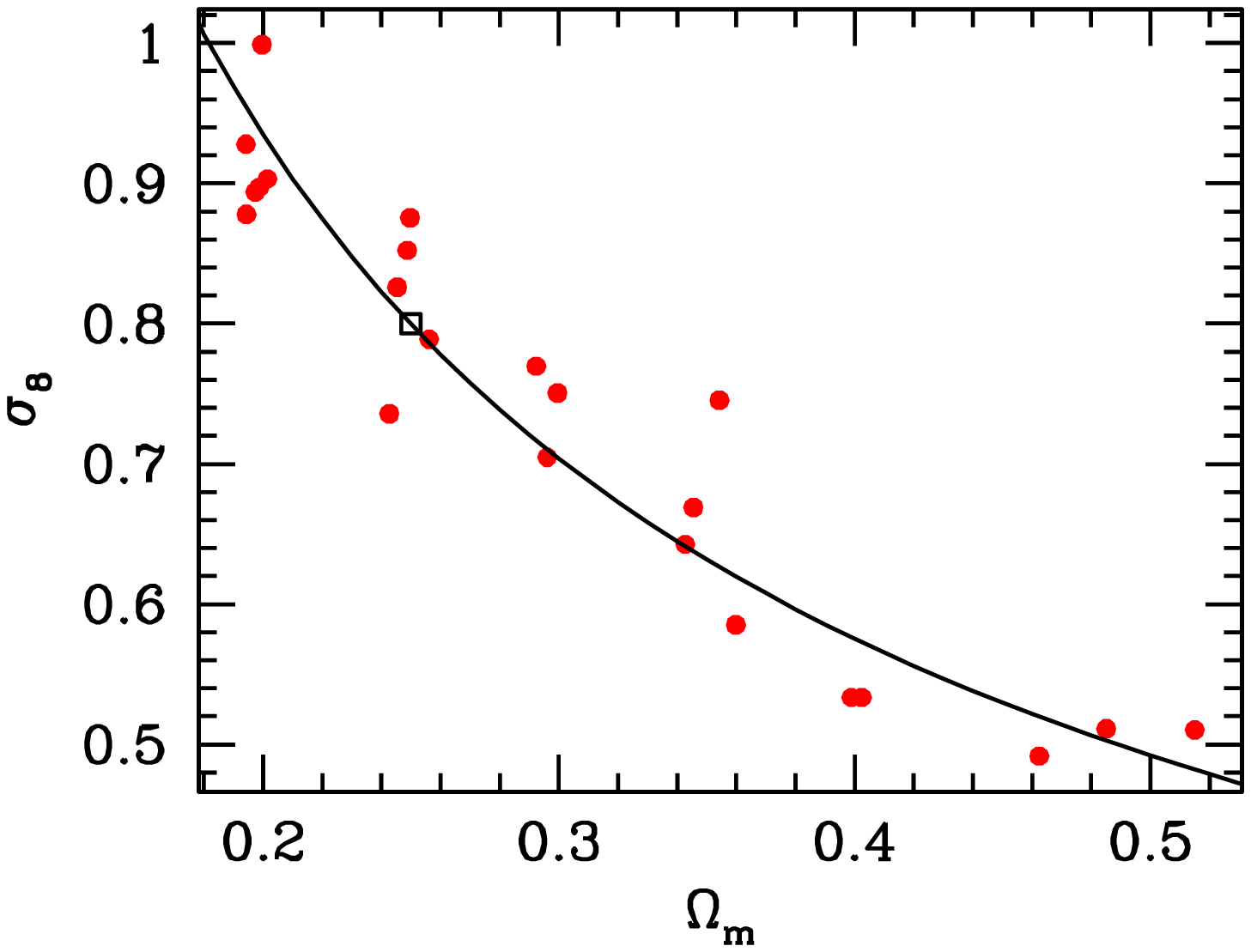}{Best fit values of the cosmological parameters extracted from the $C_l$'s of 23 mock catalogs (similar to
Fig.~\rf{bffinal} for $\xi$).}

\begin{figure*}[thbp]
\begin{center}
\sfig{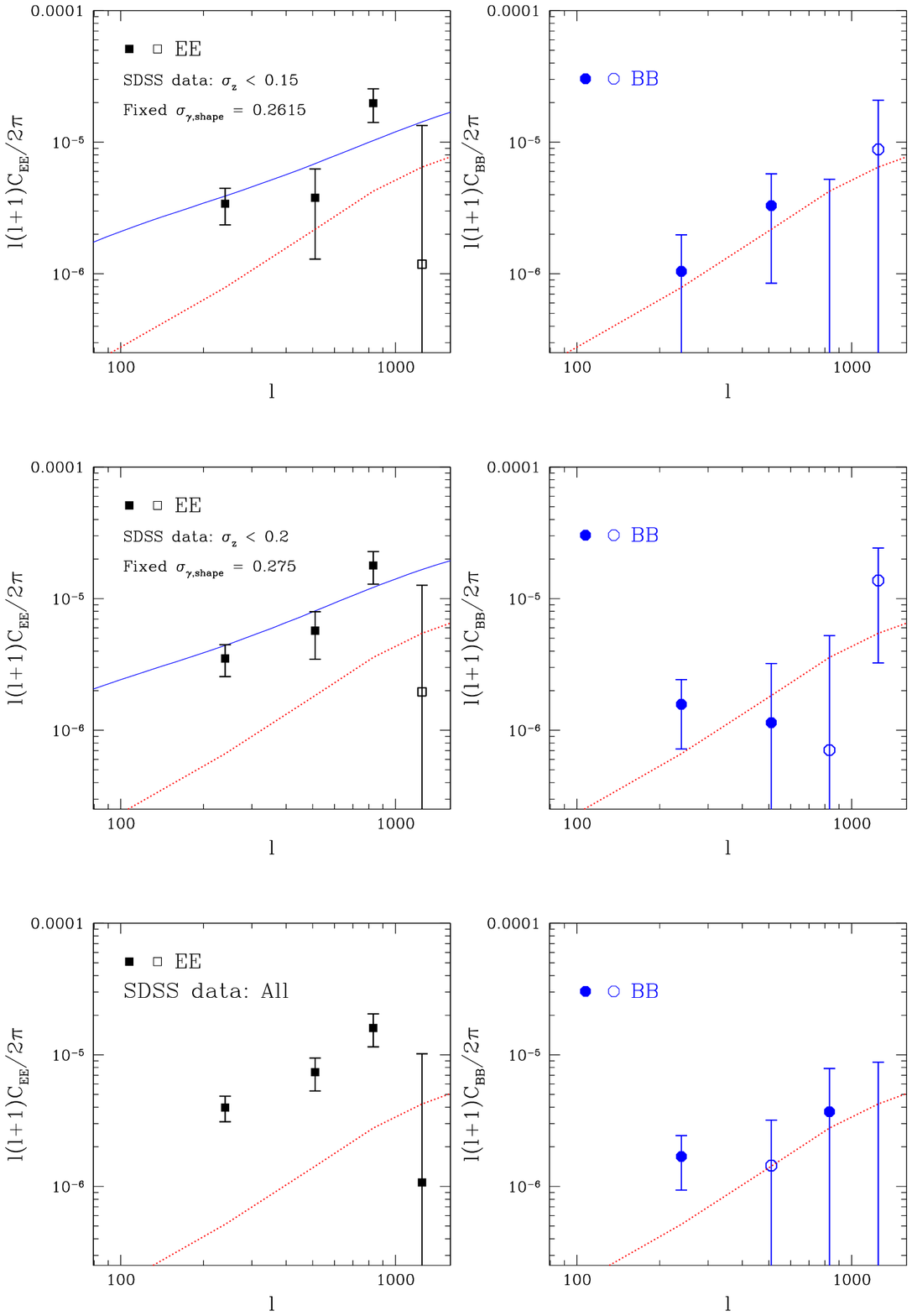}{1.6\columnwidth}
\end{center}
\caption{Estimates of the EE (left) and BB (right) power spectra for the SDSS Stripe 82 data with $\ecut[1.4]$. 
The top panels show the case with $\dzs$, and the middle panels show the case with $\dzl$. 
The bottom panels show the case with no $\sigma_z$ cut. The first two cases have quite high shape noise 
contributions. To decrease the number of degrees of freedom, we derive band powers while fixing the shape 
noise to be the measured $rms$ fluctuations in shear: $\rmstl=0.262$ and 0.275, respectively. 
The solid lines show the best-fit cosmologies for these two cases.  The red dotted lines show the level of the shape noise contribution to the error on the band power.
 As a comparison, the bottom panels for the no $\sigma_z$ cut case show the band power derived 
while the shape noise is allowed to vary. While the EE band power is derived with a higher significance, 
the uncertainty in the redshift distribution of the source galaxies make this case difficult to interpret.}\label{fig:ClEE_SDSS}
\end{figure*}

\subsubsection{Data}
We next apply this method to the real Stripe 82 data. 
Figure~\rf{ClEE_SDSS} shows the resulting QE power spectra for EE (left) and BB (right) modes for  
$\dzs$ (top panels) and $\dzl$ (middle panels), in comparison to the case with no $\sigma_z$ cut in the bottom panels (``All'' case in Figure \ref{fig:dndz}). The case with no $\sigma_z$ cut has the smallest shape noise ($\Ngal=254 /(0.1^\circ)^2$) and is the most similar to the level of shape noise in the SDSS mocks (i.e.,  $\Ngal=264 /(0.1^\circ)^2$). We detect strong EE-mode power with relatively negligible BB mode in this case, when the shape noise parameter is allowed to vary. Unfortunately, this case is not useful for deriving cosmological parameters, as we do not know the redshift distribution of the source galaxies (i.e., signal) accurately enough.

For the cases with $\dzs$ and $\dzl$, the shape noise is much larger ($129$ and $163 /(0.1^\circ)^2$, respectively). 
To decrease the degrees of freedom in accordance with the larger noise level, we either fix the BB mode to zero 
or fix the shape noise using the measured $rms$ fluctuations in shear. We adopt the latter as our main result.
Letting both vary simultaneously does not produce robust constraints with this noisy data set.

The top and the middle panels of Figure~\rf{ClEE_SDSS} show the QE power spectra when the rms shape noise 
for each component of the shear is fixed to be 0.262 and 0.275, respectively. Note that the resulting BB mode is 
consistent with zero for most of the bands. Meanwhile, the band beyond $l>1000$ is systematically low compared to the small-scale 
clustering we expect for a reasonable range of concordance cosmologies. The error bar is so large though that
including this band in the final fits does not change the parameter extraction.
For each of the measured power spectra, the solid line is the best-fit, 
flat $\LCDM$ cosmology that will be presented in Section~\ref{sec:cosmo}.
The red dotted line shows a rough estimate of the contribution of the shape noise to the error on the band power; it is close to, but smaller than the
measured power.
If we fix the BB mode to zero, instead of fixing the shape noise, we find similar results. In detail, for $\sigma_z <0.15$, the band power is slightly lower when we fix the BB mode to zero, a sign that the true shape noise level probably is slightly higher (by 1\%) than $\rmstl=0.262$ that we assumed. For $\sigma_z <0.2$, we find almost identical results for both treatments. 

\subsection{Pseudo Estimator}
We next use the pseudo estimator to derive band powers and cross-check the band power measurement derived using the quadratic estimator.

\subsubsection{Method}
Given the long and thin survey geometry of Stripe 82, one can make a flat-sky approximation and, in an ideal case, compute a discrete Fourier Transform to derive the power spectrum. A complication arises because of the incomplete coverage of the survey area and the non-uniform sampling densities. Without any treatment, the shape noise contribution in the power spectrum is derived from $\int{d\vec{x}\frac{\rmstl}{N_i}}$ rather than $\frac{\rmstl}{\bar{N}}$, where $N_i$ is the number of galaxies for the $i$th pixel and $\bar{N}$ is the average number of galaxies. It is apparent that pixels with zero galaxies will make the shape noise estimation difficult in the non-uniform sampling case. In order to resolve this issue, we design a pseudo power estimator that weighs the shear values in each pixel by the number of galaxies in that pixel, i.e., by inverse variance:
\begin{eqnarray}
\tilga \equiv \frac{N_i\gamma}{\bar{N}}.
\end{eqnarray}
Then this makes the shape noise contribution to the power spectrum be simply $\frac{\rmstl}{\bar{N}}$. 

Meanwhile, such a weighting results in a convolution of power in Fourier space and therefore in mode-mixing between E- and B-modes. That is, the estimator is biased relative to the true power spectrum and also it is not an optimal estimator. In order to derive an unbiased estimate of the true power spectrum from the pseudo power spectrum, we need to construct the mode-coupling matrix based on the number weighting scheme we used and deconvolve it by matrix inversion. \citet{Hikage:2010sq} have shown the procedure for deriving the unbiased estimate of the true power spectrum from the pseudo power spectrum in the presence of a survey masking effect. Our case is analogous to theirs except that our pseudo power spectrum includes the effect of the number weighting scheme rather than the masking effect. We therefore follow the deconvolution procedure in \citet{Hikage:2010sq} to remove the mode-coupling effect due to the weighting as well as the finite sky effect. We refer readers to \citet{Hikage:2010sq} for further details for setting up these matrix operations.

\subsection{Pseudo Estimator vs. Quadratic Estimator}
In Figure~\rf{sdssmockpe}, we show the average and standard deviation of 23 SDSS mock power spectra derived using the pseudo estimator (red triangles for E-mode and magenta pentagons for B-mode) in comparison with the quadratic estimator (black squares for E and blue circles for B-mode). For the pseudo estimator, we use the input $rms$ shape noise to subtract off the shape noise contribution. We cut off $l>1000$ for the pseudo estimator: our pixel resolution causes a non-zero aliasing of power from smaller scales onto this scale. We observe a sign of deviation from the input power spectrum in the E-mode of the pseudo estimator at $l\sim 160$, but overall the results between the pseudo estimator and the quadratic estimator appear quite consistent. 
\begin{figure}
\sfig{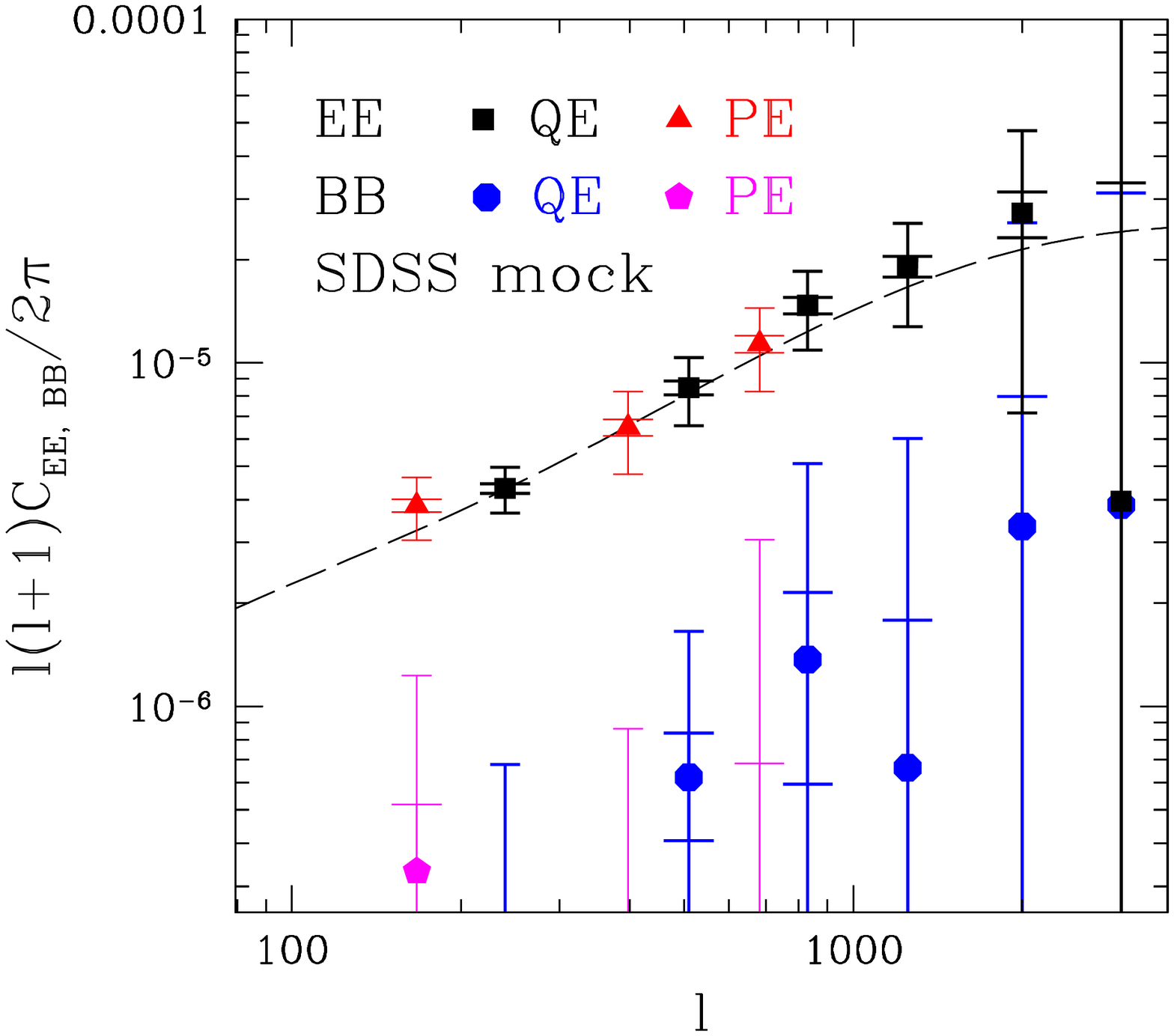}{1.0\columnwidth}
\caption{{\small Average power spectra of the 23 SDSS mocks using the pseudo estimator (PE) in comparison with the quadratic estimator (QE). There are two sets of error bars. The larger set of error bars show the standard deviation among the 23 mocks. The smaller set of error bars show the error associated with the average of the 23 mocks, i.e., the standard deviation divided by $\sqrt{23}$. The dashed line shows the input EE power spectrum. The input BB power is zero. }}\label{fig:sdssmockpe}
\end{figure}

We next apply the pseudo estimator to the Stripe 82 data. We again assume $\rmstl=0.262$ and 0.275, respectively, for each of the two data sets with $\sigma_z < 0.15$ and $\sigma_z < 0.2$. The error bars are derived by regenerating SDSS mocks using the sampling density distribution and $\rmstl$ of each of the two data sets and by taking the dispersion among the 23 mocks.  Figure~\rf{sdsspeqe} shows the resulting band power of the Stripe 82 data using the pseudo estimator, in comparison to the quadratic estimator. Overall we find consistency between the two estimators. 

\begin{figure*}
\sfig{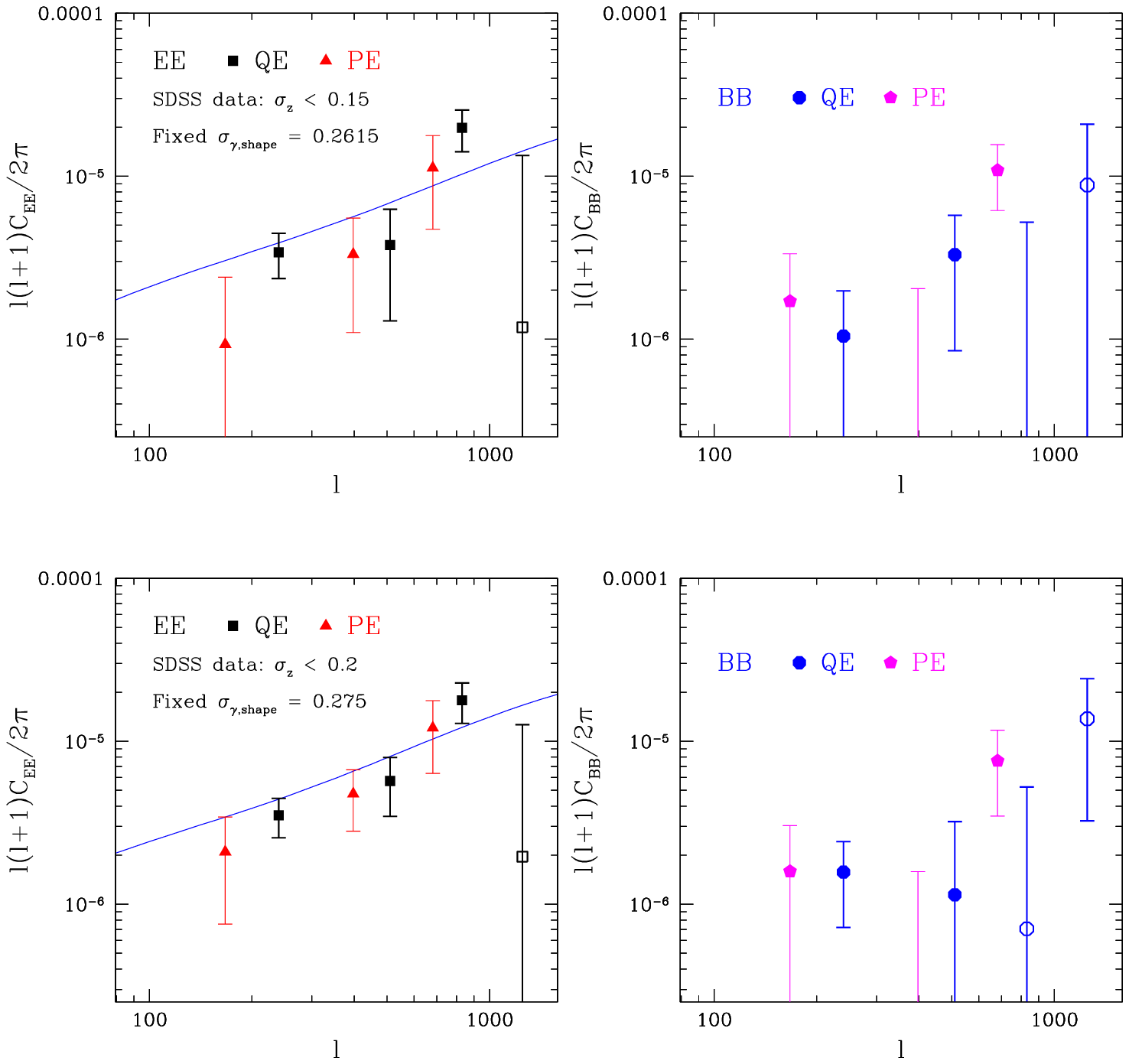}{2.0\columnwidth}
\caption{{\small Estimates of the pseudo EE (left) and BB (right) power spectra for the SDSS Stripe 82 data (red triangles for E-mode and magenta pentagons for B-mode). We also plot the results of the quadratic estimator for comparison (black squares for E and blue circles for B-mode). The solid blue line shows the best-fit cosmology to the quadratic estimator. We find an overall consistency between the two estimators. }}
\label{fig:sdsspeqe}
\end{figure*}

\section{Constraints on Cosmological parameters}\label{sec:cosmo}

The cosmic shear measurements described above are most sensitive to the 
power on scales of $10-60'$ or $l\sim100-800$. This power is sensitive 
primarily to the amplitude of the fluctuations $\sigma_8$ and the matter 
density $\Omega_m$; Fig.~\rf{powerl} displays the sensitivity to $\sigma_8$ for fixed $\Omega_m$.

\Sfig{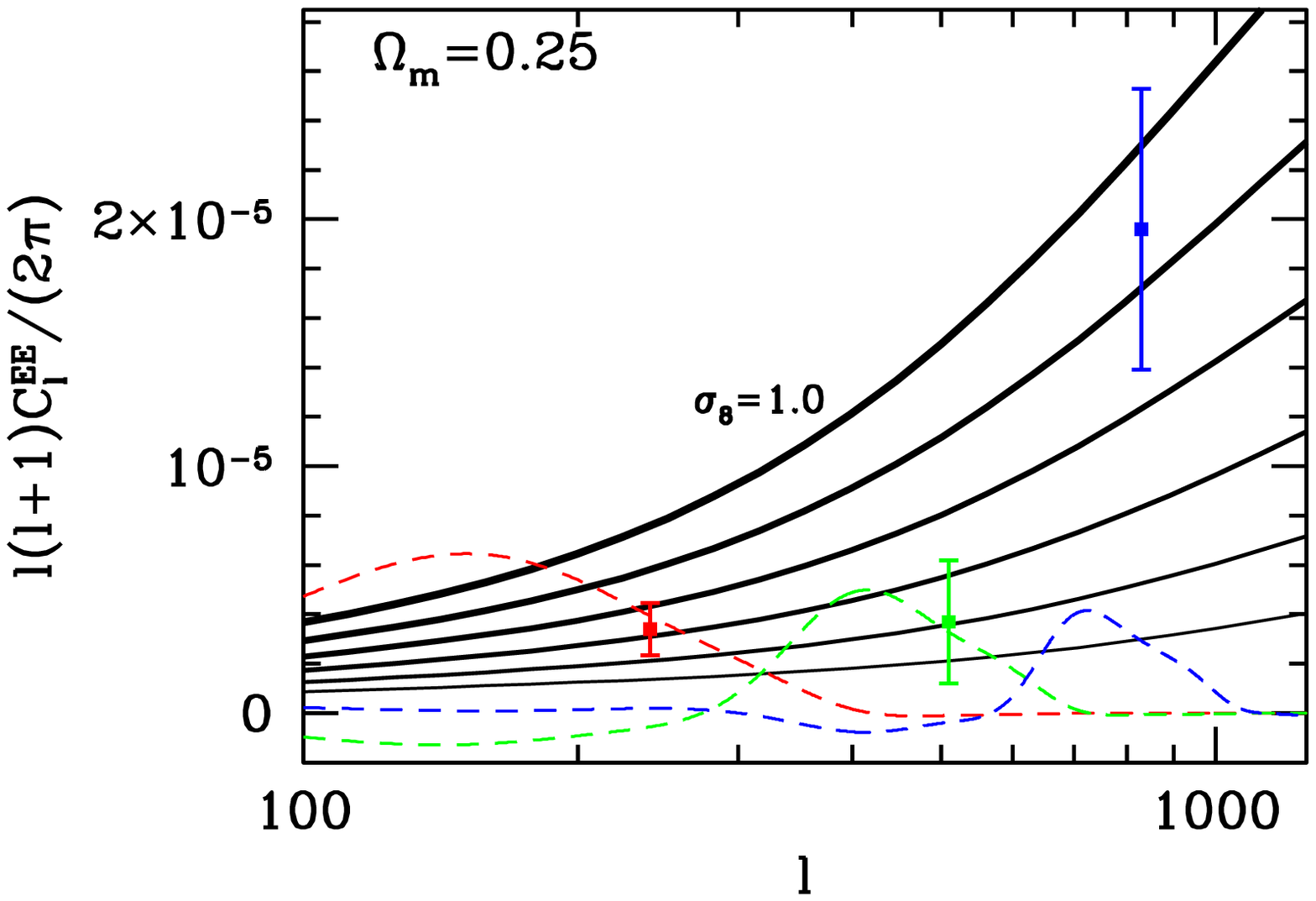}{The power spectrum of the convergence, the $EE$ spectrum, for different values of $\sigma_8$ 
and fixed $\Omega_m$ 
together with the data from \S\ref{sec:power} ($\sigma_z<0.15$). Each (colored) data point is shown 
at the central value of its $l$-band. The result, though, can be interpreted only with the aid of the window 
function (dashed curves; unnormalized here), so, e.g., the band centered at $l=240$ is actually most 
sensitive to power at $l\sim170$. }

The theoretical predictions are obtained by convolving the nonlinear power spectrum in \ec{pkappa} 
with the window function describing the galaxy distribution, as plotted in Fig.~\rf{w}. We approximate 
the nonlinear spectrum as a function of $k$ and $z$
using {\tt halofit}~\citep{Smith:2002dz}. Recent work~\citep{Eifler:2010kt} has shown that 
{\tt halofit} underpredicts the power spectrum by 6-7\% on the scales of interest, compared 
with accurately calibrated simulations~\citep{Lawrence:2009uk}. Our implementation of {\tt halofit}
also differs from the simulations at about the same level at $z=0$, but fortuitously agrees with them
to within a few percent
on the scales and redshifts probed by the Stripe 82 data. The systematic error on the cosmological parameters
due to this theoretical uncertainty
then should be at most $3\%$ (since the power spectrum scales roughly as the square of $\sigma_8$) and is probably far
below that. It is much smaller than the statistical uncertainty. 

A final check is to compare the constraints in parameter space from the different data sets ($\sigma_z<0.15, 0.20$) 
and techniques ($C_l, \xi_+, \xi_{E}$). It is simplest to compare the constraints on a single parameter, 
instead of in the 2D ($\sigma_8,\Omega_m$) plane. As shown in Fig.~\rf{bffinal}, the data are most sensitive 
to the combination $\Omega_m^{0.7}\sigma_8$; Fig.~\rf{limits} shows the constraints on this combination 
for the different ways of analyzing the data. The different cuts and techniques lead to consistent constraints. 
We also note that our best-fit $\chi^2 = 6$ using the $\xi$ data for
the $\sigma_z < 0.15$ sample, while
$\chi^2 = 35$ assuming no signal ($\xi=0$); the resulting $\Delta\chi^2=29$,
indicating 5-$\sigma$ detection of the cosmic shear signal in our data.

\Sfig{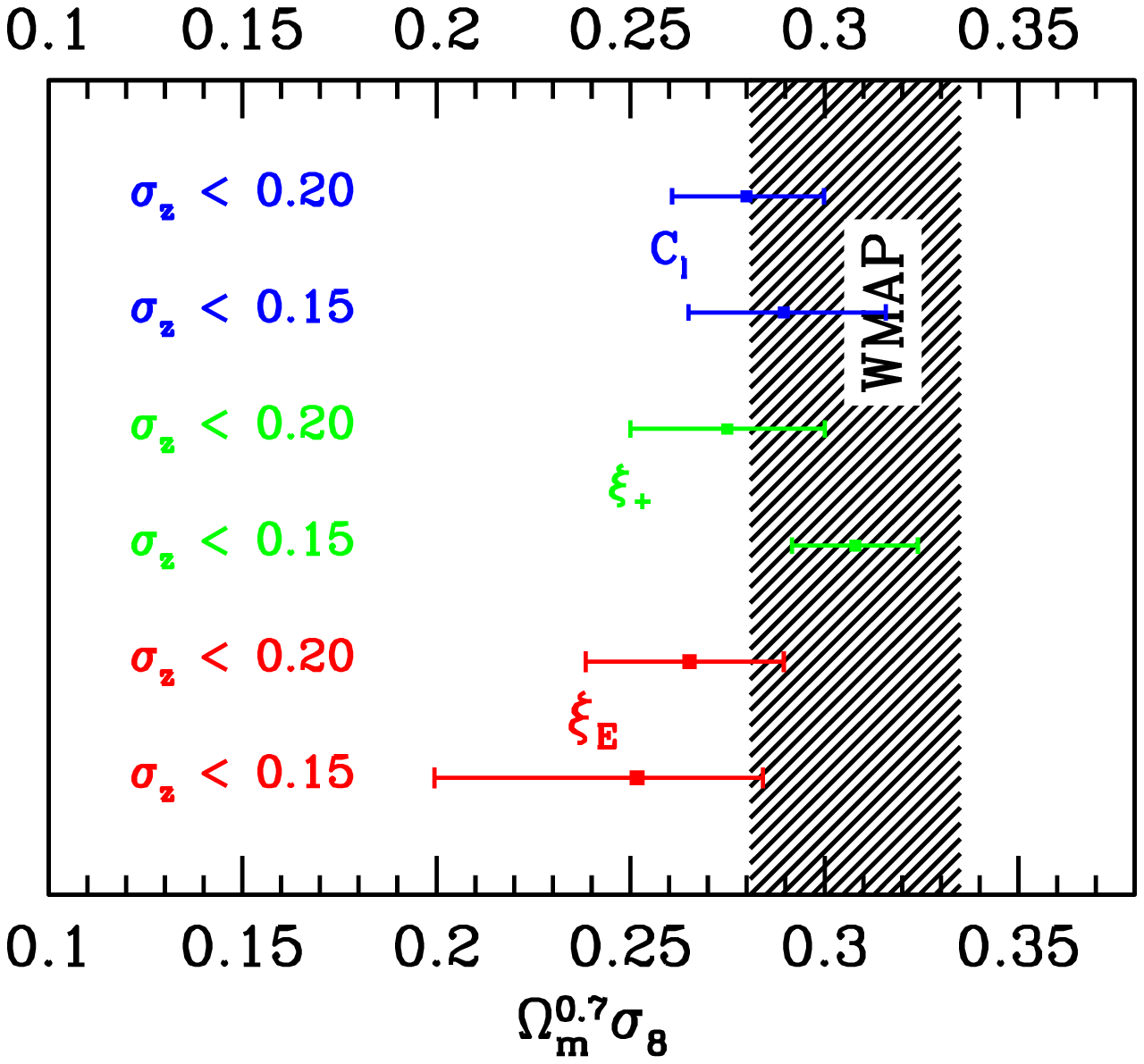}{$68\%$ Confidence levels on the parameter $\Omega_m^{0.7}\sigma_8$ from the different 
data sets and analysis techniques. Also shown is the 1-$\sigma$ band from WMAP~\citep{Larson:2010gs}.}

For our final constraints, we choose the most conservative cut, the data set with $\sigma_z<0.15$, 
using $\xi_E$, which is un-contaminated by the B-modes. Fig.~\rf{chifinalwmap} shows our results in 
the $\Omega_m,\sigma_8$ plane. The 1-$\sigma$ range on the constrained parameter is 
\be 
\Omega_m^{0.7}\sigma_8 = 0.252^{+0.032}_{-0.052}.
\ee
The error bars are consistent with those obtained with the mock catalogs, accounting for the lower galaxy density
in the sample with the photo-$z$ cuts.

\Sfig{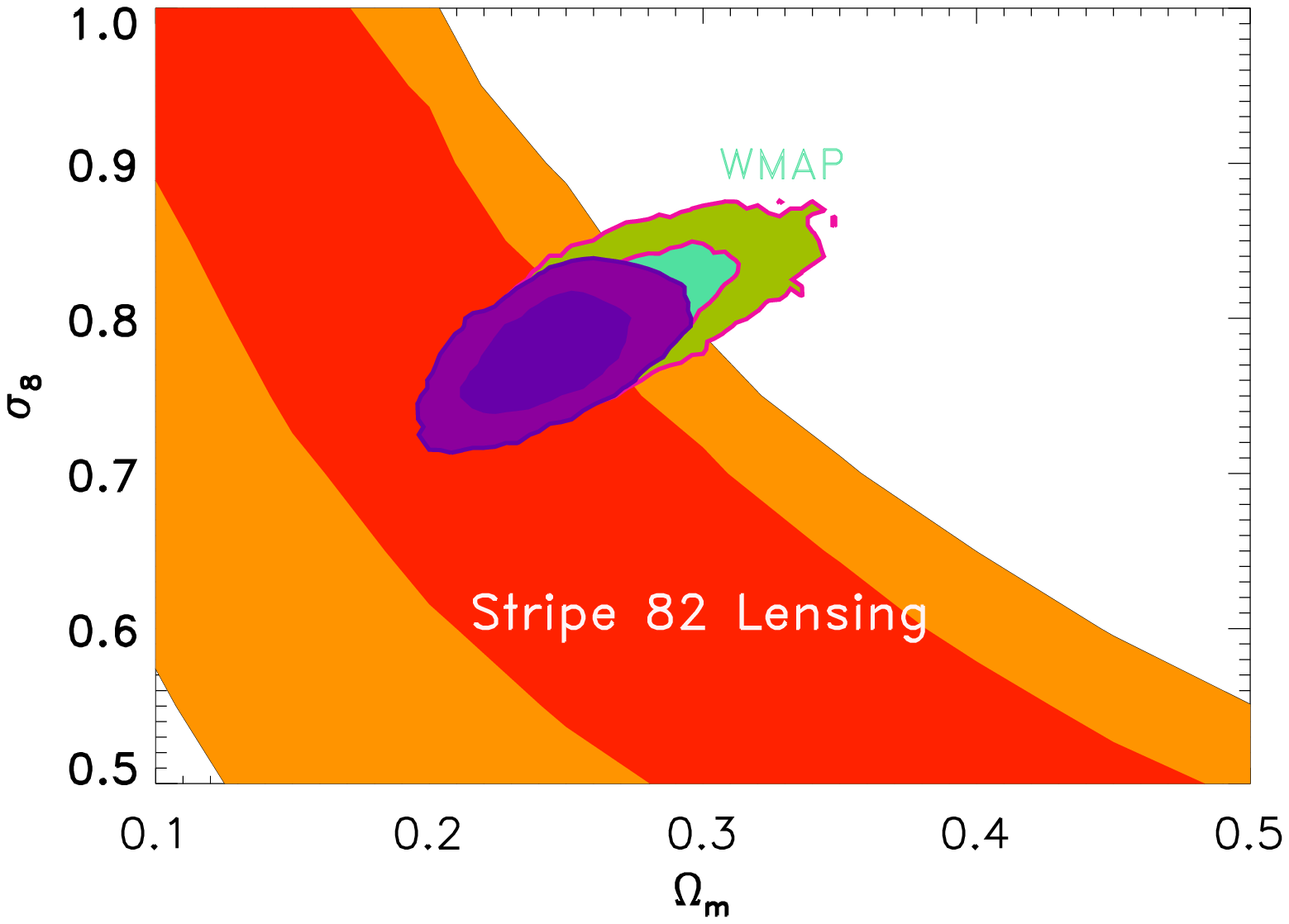}{Constraints on cosmological 
parameters from SDSS Stripe 82 cosmic shear at the 1- and 2-$\sigma$
level. Also shown are the constraints from WMAP. The innermost region is the 
combined constraint from both WMAP and
Stripe 82.}

Fig.~\rf{chifinalwmap} shows the constraints in the 2D plane along with the constraints from the WMAP 7-year
data \citep{Komatsu:2010}. The two data sets give consistent results, and the complementarity tightens the constraints on
$\Omega_m$ at the high end.

\section{Conclusions}\label{sec:conclusions}
We have performed a cosmic shear analysis of the SDSS coadd 
data \citep{Annis:2011}, 
a 275 square degree area of the SDSS imaging data where we achieve 
2 magnitudes fainter than the nominal depth of the survey through the 
coaddition of multiple exposures. Photometric redshifts for the 
coadd galaxies were obtained from their colors using a neural network
algorithm \citep{Reis:2011} and corrections to the PSF modeling for 
accurate galaxy shape measurements were
implemented as part of this work (Section~\ref{sec:data}).

Through a  $>5\sigma$ detection of the cosmic shear signal in
the SDSS coadd data, we have 
measured the combination of the matter density $\Omega_m$ and the amplitude of 
matter fluctuations $\sigma_8$ in the Universe. 
The measurement was performed using both the shear-shear
angular correlation function in real space (Section~\ref{sec:xi}) and
the power spectrum in Fourier space (Section~\ref{sec:power}). 
We tested the quadratic estimator introduced by \cite{Hu:2000ax}  
for the power spectrum, plus our own version of the pseudo estimator;
such power spectrum methods have not been commonly used in
previous cosmic shear analyses of real data.
Our power spectrum results are consistent with each other and with
our correlation function results (Fig.~\ref{fig:limits}).  
Our cosmological parameter constraint may be expressed as
$\Omega_m^{0.7}\sigma_8 = 0.252^{+0.032}_{-0.052}$, which is
in good agreement with WMAP (Fig.~\ref{fig:chifinalwmap}),
as well as with other recent weak lensing surveys, in particular
the next two largest area samples, CFHTLS \citep[][at 57 deg$^2$]{Fu:2007qq}
and CTIO \citep[][at 75 deg$^2$]{Jarvis:2003}, plus the much deeper 
COSMOS data set \citep{Schrabback:2010}.

We have shown that the systematic effects on the correlation 
function due to PSF mis-modeling in our data  are at the sub-percent level for 
scales up to 2$\arcdeg$. 
We also tested samples with different photometric redshift error cuts. 
The  most conservative $\sigma_z$ cuts result in 
slightly better agreement among the different methods, 
indicating a minor effect and confirming the assumption that accurate photo-$z$
estimates are crucial for cosmic shear studies.   

Our 275~deg$^2$ Stripe 82 coadd data is the largest area survey for which
cosmic shear has been measured, and our results pose an important 
precedent for future analyses of even larger area surveys, such as the 
Dark Energy Survey and LSST.


\clearpage

\acknowledgements
We thank Chiaki Hikage for his essential help during the implementation of the pseudo estimator. 
This work is supported by the US Department of Energy, including grant
DE-FG02-95ER40896 and by the National Science Foundation under Grant AST-0908072.
Some of this work was performed on the Joint Fermilab-KICP Supercomputing Cluster, supported by  Fermilab, Kavli Institute for Cosmological Physics and the University of Chicago. 
Fermilab is operated by Fermi Research Alliance, LLC under Contract
No. DE-AC02-07CH11359 with the United States Department of Energy. 

Funding for the SDSS and SDSS-II has been provided by the Alfred P. Sloan Foundation, the Participating Institutions, the National Science Foundation, the U.S. Department of Energy, the National Aeronautics and Space Administration, the Japanese Monbukagakusho, the Max Planck Society, and the Higher Education Funding Council for England. The SDSS Web Site is http://www.sdss.org/.

The SDSS is managed by the Astrophysical Research Consortium for the Participating Institutions. The Participating Institutions are the American Museum of Natural History, Astrophysical Institute Potsdam, University of Basel, University of Cambridge, Case Western Reserve University, University of Chicago, Drexel University, Fermilab, the Institute for Advanced Study, the Japan Participation Group, Johns Hopkins University, the Joint Institute for Nuclear Astrophysics, the Kavli Institute for Particle Astrophysics and Cosmology, the Korean Scientist Group, the Chinese Academy of Sciences (LAMOST), Los Alamos National Laboratory, the Max-Planck-Institute for Astronomy (MPIA), the Max-Planck-Institute for Astrophysics (MPA), New Mexico State University, Ohio State University, University of Pittsburgh, University of Portsmouth, Princeton University, the United States Naval Observatory, and the University of Washington.

\bibliography{lensing,bib}

\end{document}